\documentclass[11pt]{llncs}
\usepackage[margin=1in]{geometry}

\usepackage{preamble}
\usepackage{maths}

\def\Ball{\mathrm{Ball}}

\allowdisplaybreaks %to allow page break in equations

\title{Improved Bounds for Online Facility Location with Predictions%
\thanks{This work has been partially supported by project MIS 5154714 of the National Recovery and Resilience Plan Greece 2.0 funded by the European Union under the NextGenerationEU Program. A significant part of this work was done while Nikolas Patris was an undergraduate student at the National Technical University of Athens.}}

\author{Dimitris Fotakis\inst{1,2} \and Evangelia Gergatsouli\inst{3} \and Themis Gouleakis\inst{4} \and \\ Nikolas Patris\inst{5} \and Thanos Tolias\inst{1,2}}

\institute{%
National Technical University of Athens, Greece
\and
Archimedes/Athena RC, Greece
\and 
University of Wisconsin - Madison, WI, USA
\and 
National University of Singapore, Singapore
\and
University of California, Irvine, CA, USA\\[2pt]
\email{fotakis@cs.ntua.gr}, \email{gergatsouli@wisc.edu},
\email{themis.gouleakis@gmail.com}, \email{npatris@uci.edu}, \email{thanostolias@gmail.com}}

\begin{document}
\maketitle

\begin{abstract}
We consider Online Facility Location in the framework of learning-augmented online algorithms. In Online Facility Location (OFL), demands arrive one-by-one in a metric space and must be (irrevocably) assigned to an open facility upon arrival, without any knowledge about future demands. We focus on uniform facility opening costs and present an online algorithm for OFL that exploits potentially imperfect predictions on the locations of the optimal facilities. We prove that the competitive ratio decreases from sublogarithmic in the number of demands $n$ to constant as the so-called $\eta_1$ error, i.e., the sum of distances of the predicted locations to the optimal facility locations, decreases. E.g., our analysis implies that if for some $\eps > 0$, $\eta_1 = \opt / n^\eps$, where $\opt$ is the cost of the optimal solution, the competitive ratio becomes $O(1/\eps)$. We complement our analysis with a matching lower bound establishing that the dependence of the algorithm's competitive ratio on the $\eta_1$ error is optimal, up to constant factors. Finally, we evaluate our algorithm on real world data and compare the performance of our learning-augmented approach against the performance of the best known algorithm for OFL without predictions. 
\end{abstract}

%%%%%%%%%%%%%%%%%%%%%%%%%%%%%%%%%%%%%%%%%%%%%%%%%%%%%%%%%%%%%%%
\pagenumbering{arabic}
\pagestyle{plain}

\section{Introduction}
\label{sec:intro}

Online algorithms is a field that deals with decision making in cases where the input data is not known in advance, but rather arrives in a sequential way. The algorithm is required to make irrevocable decisions, only based on the input data received at a given point in time, and to incur the corresponding irrevocable cost for each of them. Traditionally, in the analysis of online algorithms we assume, rather pessimistically, that an adversary always presents the algorithm with the worst case input. More precisely, the performance of online algorithms is evaluated by the \emph{competitive ratio} \cite{BoroYavi1998}, which is the worst-case ratio of the total algorithm's cost to the cost of a computationally
unrestricted optimal algorithm aware of the entire request sequence in advance. 

On the other hand, machine learning (ML) aims to \emph{predict} the unknown based on historical data and to \emph{learn} how the world looks like, rather than dealing with worst-case scenarios. A recent trend aims to use machine learning predictions about the future input in order to deal with the inherent uncertainty in online algorithms, while still providing worst case performance guarantees. Specifically, one might think that directly using machine learning in online problems should enhance their performance, since by predicting the input, with some error, we should be able to come up with almost optimal solutions. In reality, this turns out not to be true, since the error of the learner does not necessarily remain constant and could propagate along different phases of the algorithm. %and result in a much larger error.

Lykouris and Vassilvitski \cite{LykoVass2018} proposed a framework aiming to provide formal guarantees for such \emph{learning-augmented} online algorithms, in terms of their \emph{consistency} and \emph{robustness}. They require the algorithm to be near optimal, if the predictions about the future input are accurate (consistency), while for arbitrary erroneous predictions, the competitive ratio should gracefully degrade to (and not exceed by far) the worst-case one (robustness). Generally, the idea of combining online algorithms with ML advice is that in the end, we should be able to overcome the traditional worst-case lower bounds and get the best of both worlds. The learning-augmented algorithm, given some predictions of total error $\eta$, is required to make decisions online. In the end, following the ideas of \cite{LykoVass2018,PuroSvitKuma2018,AntoGoulKleeKole2020}, the competitive ratio is given as a function of $\eta$.
Many online problems have already been studied under this framework, such as ski rental, scheduling, the secretary problem, metrical task systems (MTS) and more (see e.g., the survey of \cite{MitzVas20}). %, for which we have included a brief overview in the related work section. 
In this work we investigate the competitive ratio of Online Facility Location in the framework of learning-augmented online algorithms, following an approach similar to \cite{AntoGoulKleeKole2020,JiangLLTZ22,LykoVass2018,PuroSvitKuma2018}.

% \paragraph{Online Facility Location.}
\smallskip\noindent{\bf Online Facility Location.} 
In Online Facility Location (OFL), introduced by Meyerson \cite{Meye2001}, we are presented with a sequence of demands located in an underlying metric space. Each demand must be connected to an open facility upon arrival. Each facility has an opening cost, which is irrevocable in the sense that once opened, facilities cannot be closed. Every demand incurs its assignment cost, which is the distance to the closest open facility at the demand's assignment time. Our goal is to decide where to open the facilities and where to assign each arriving demand, while incurring the minimum possible facility opening plus demand assignment cost. Meyerson presented an elegant randomized algorithm with competitive ratio $O(\log n / \log \log n)$, where $n$ is
the number of demands. After Meyerson's initial result, there has been a significant volume of work on OFL and its many variants. We provide a brief list in the related work section.

\smallskip\noindent{\bf Online Facility Location with Predictions.} 
In OFL with predictions (OFLpred), every demand $v$ in the request sequence is accompanied with a prediction $p$ on the location of the optimal facility where $v$ is assigned. Every demand must be connected to an open facility upon arrival, and facility and assignment costs are irrevocable and are defined as in standard OFL. In addition to the number $n$ of demand-prediction pairs, instances of OFLpred are parameterized by the \emph{total prediction error} $\eta_1$, which is the sum, over all predictions, of their distance to the respective optimal facility, and the \emph{maximum prediction error} $\eta_\infty$, which is the maximum such distance (and hence $\eta_\infty \leq \eta_1)$. %A natural question is to which extent the (possibly erroneous) predictions could contribute towards an improved competitive ratio. 

H.-C.~Jiang et al. \cite{JiangLLTZ22} were the first to study the competitive ratio of OFLpred as a function of $\eta_\infty$. They considered non-uniform facility costs, carefully adapted Meyerson's algorithm, and proved that its competitive ratio is $O(\log \min\{ \frac{n \eta_\infty}{\opt}, n\})$, where $\opt$ is the optimal cost. They also showed an almost matching lower bound of  $\Omega\!\left(\frac{\log \min\{ \frac{n \eta_\infty}{\opt}, n \}}{\log\log n}\right)$ on the competitive ratio of any randomized algorithm for OFLpred with uniform facility costs. The results of \cite{JiangLLTZ22} imply that the competitive ratio of OFLpred degrades from a small constant, if all predictions are perfect (consistency), to logarithmic, if there are some inaccurate predictions and $\eta_\infty \geq \opt$ (robustness). Moreover, for the class of instances in the lower bound of \cite{JiangLLTZ22}, where a small fraction of the predictions have error $\eta_\infty$ and the remaining ones are collocated with the respective demand points (and thus, $\eta_1 = \Omega(\opt)$ as long as $\eta_\infty \geq \opt / n^{1-\delta}$, for any constant $\delta > 0$), an almost logarithmic competitive ratio is unavoidable.

\subsection{Motivation and Contribution}
\label{subsec:contribution}

Our work is motivated from the observation that determining the dependence of OFLpred's competitive ratio on the total prediction error $\eta_1$ (in addition to its dependence on $\eta_\infty$ studied in \cite{JiangLLTZ22}) contributes to a deeper understanding about how and to which extent predictions can help in improving the performance of OFL algorithms. 
%
%\cite{JiangLLTZ22} mostly ignore the dependence of OFLpred's competitive ratio on $\eta_1$. 
%
E.g., let us consider OFLpred instances with $\eta_\infty = \opt / n^\eps$, for some constant $\eps > 0$. Within this class, the upper and the lower bound of \cite{JiangLLTZ22} fail to differentiate, as far as their best possible competitive ratio is concerned, between (i) instances where all but few predictions are perfect; (ii) instances where for some $\beta \in (0, 1)$, every prediction is $1/\beta$ times closer to the respective optimal facility than the corresponding demand; and (iii) instances where every prediction is at distance $\eta_\infty$ to the optimal facility. In (i), $\eta_1 \approx \opt / n^\eps$ and we should expect a constant competitive ratio, if $\eps$ is a constant. In (ii), $\eta_1 \approx \beta \opt$ and the competitive ratio must be an increasing function of $\beta$. Only in (iii), we should expect a competitive ratio fully determined by $\eta_\infty$. For all these instances with very different prediction quality (and for many other instances in between), the results of \cite{JiangLLTZ22} guarantee a competitive ratio of $O((1-\eps) \log n)$, while their lower bound applies only to (iii). For another example, if $\eta_\infty = \opt / (\log n)^\ell$, for some integer $\ell \geq 1$, the resulting guarantee on the competitive ratio is $O(\log n - \ell\log\log n)$ in all cases (i)-(iii). 

In this work, we focus on OFLpred with \emph{uniform} facility costs, where
the cost of opening a facility at any point of the underlying metric space is $f$, and determine OFLpred's competitive ratio as a function of both $\eta_\infty$ and $\eta_1$. As demonstrated by the above discussion we believe that the total error $\eta_1$ is at least as representative of the prediction's accuracy as the maximum error $\eta_\infty$. 

In Section~\ref{sec:algo}, we present a randomized algorithm that works similarly to Meyerson’s algorithm \cite{Meye2001}, but with the predictions in place of the demands. Specifically, our algorithm decides on whether to open a facility at the predicted location with probability proportional to the distance of the predicted location to the nearest open facility. By generalizing the analysis of \cite{Meye2001}, so that it also takes the prediction error of each demand-prediction pair into account, we show that if all predictions are perfect, our algorithm is $2$-competitive ($2$-consistency), and that for any error $\eta_1$ and $\eta_\infty$ with $\opt \geq \eta_1 \geq \eta_\infty > \opt / n$, the competitive ratio of our algorithm is $O\!\left(\frac{\log\!\left(\frac{n \eta_\infty}{\opt}\right)}{\log\!\left(\frac{\opt}{\eta_1}\log\!\left(\frac{n \eta_\infty}{\opt}\right)\right)}\right)$.

In Section~\ref{sec:combination}, we show how to apply cost doubling to OFLpred in order to obtain an online algorithm with thrice the minimum competitive ratio of our proposed algorithm and the algorithm of \cite{JiangLLTZ22}. For uniform facility costs, the latter works as Meyerson’s algorithm, but it opens facilities in pairs, one at the demand's location and another at the respective predicted optimal location. Our generalized analysis framework gives an improved competitive ratio for \cite{JiangLLTZ22}'s algorithm with uniform facility costs. Taking the minimum of the competitive ratios of the two algorithms, we get: 

\begin{theorem}[Upper Bound]\label{thm:main_intro}
The competitive ratio of OFLpred is at most: 
\[
\min\left\{ 
\begin{array}{ll}
9\frac{n \eta_\infty}{\opt} + 6 
& \mbox{for all $\eta_\infty \geq 0$}
\\[2pt]
O\!\left(\frac{\log\min\left\{ \frac{n \eta_\infty}{\opt}, n\right\} }{\log\left(\max\left\{ \frac{\opt}{\eta_1}, 1 \right\} \log\min\left\{ \frac{n \eta_\infty}{\opt}, n\right\}\right)}\right)
\ \ & 
\mbox{for all $\eta_1 \geq \eta_\infty > \frac{\opt}{n}$}
\end{array}\right.\]
%
%where $n$ is the number of demand points and $\opt$ is the cost of an optimal solution. 
\end{theorem}

We note that $\eta_1$ and $\eta_\infty$ are only used in order to bound the competitive ratio of our algorithm, since its execution does not require their knowledge. Moreover, our algorithm is robust in the sense that if $\eta_1 \geq \opt$ and $\eta_\infty \geq \opt$, the algorithm's worst-case competitive ratio is $O(\log n / \log \log n)$. It is interesting that the competitive ratio of Theorem~\ref{thm:main_intro} (which is best possible, as shown by Theorem~\ref{thm:lower_intro} below) is achieved by combining two versions of Meyerson's algorithm: ours, which uses the predicted locations for its facilities and decisions, and \cite{JiangLLTZ22}'s, whose decisions are guided by the demand locations and lets the predicted locations play a supporting role (see also \cite{Wei2020} for a similar construction in the context of online caching with predictions). 

In Section~\ref{sec:lower_bound}, we prove a lower bound on the competitive ratio of OFLpred, establishing that the dependence of the competitive ratio of Theorem~\ref{thm:main_intro} on $\eta_1$ and $\eta_\infty$ is best possible. The lower bound construction generalizes the lower bounds of \cite{Fota2008} and \cite{JiangLLTZ22} and requires non-trivial modifications of the underlying metric space and the sequence of demand-prediction pairs so that we can achieve virtually any allowable combination of maximum $\eta_\infty$ and total $\eta_1$  error (see also Remark~\ref{rem:lower_bound}, Section~\ref{sec:lower_bound}). 

\begin{theorem}[Lower Bound]\label{thm:lower_intro}
For all $\alpha = \eta_\infty / \opt < 1/3$ and $\beta \approx \eta_1 / \opt \in (3\alpha, 1]$, there are OFL instances with $n$ demand-prediction pairs where any randomized algorithm has competitive ratio 
\[
\Omega\!\lp( \frac{\log(\alpha n) }{\log(\log(\alpha n)/ \beta) } \rp) 
\]
\end{theorem}

%\smallskip\noindent{\bf Comparison with \cite{JiangLLTZ22}.}
%
Although our techniques are quite different from those in \cite{JiangLLTZ22}, from a conceptual viewpoint, theorems~\ref{thm:main_intro}~and~\ref{thm:lower_intro} significantly generalize and can be regarded as an informative refinement of the results in \cite{JiangLLTZ22}. Specifically, for any fixed $\alpha = \eta_\infty / \opt$, theorems~\ref{thm:main_intro}~and~\ref{thm:lower_intro} determine the best possible competitive ratio of OFLpred as a function of $\beta = \eta_1 / \opt$. Returning to our motivating example at the beginning of this section, 
we can now differentiate between cases (i)-(iii). Theorems~\ref{thm:main_intro}~and~\ref{thm:lower_intro} imply that the best possible competitive ratio is $\Theta(1/\eps)$ in (i), $\Theta(\frac{(1-\eps)\log n}{\log((1-\eps)\log( n)/\beta)})$ in (ii), and $\Theta(\frac{(1-\eps)\log n}{\log((1-\eps)\log n)})$ in (iii). Similarly, if for some integer $\ell \geq 1$, $\eta_1 \approx \eta_\infty = \opt / (\log n)^\ell$, we prove that the best possible competitive ratio for case (i) is $\Theta(\frac{\log n}{(\ell+1)\log\log n)})$. 

Finally, in Section~\ref{sec:experiments}, we experimentally evaluate our proposed algorithm on both real-world and synthetic datasets for different types of predictions, with different prediction error $\eta_1$ and $\eta_\infty$.

\subsection{Other Related Work}
\label{subsec:related}

\noindent\textbf{Learning Augmented Algorithms.}
This line of work was initiated by Munoz and Vassilvitski \cite{MunoVass2017} and Lykouris and Vassilvitski \cite{LykoVass2018}, who formally introduced the notions of consistency and robustness. Purohit et al.  \cite{PuroSvitKuma2018} considered ski rental and non-clairvoyant scheduling, giving consistency and robustness guarantees that depend on a hyperparameter that has to be given to the algorithm in advance. 
%
%Following this, Gollapudi and Panigrahi \cite{GollPani2019} also considered the ski rental problem in the setting of multiple predictors, while Wang et al.~\cite{WangLiWang2020} considered the multi-shop ski-rental problem, a generalization of the classic ski rental problem.  
%
Lykouris and Vassilvitskii \cite{LykoVass2018} studied the classical online caching problem and were able to adapt the Marker algorithm \cite{FiatKarpLubyMcgeSleaYoun1991} to obtain a tradeoff between robustness
and consistency. Rohatgi \cite{Roha2020} and Wei \cite{Wei2020} subsequently presented simpler learning-augmented caching algorithms with improved dependence of their competitive ratios on the prediction errors. Interestingly, Wei's algorithm boils down to selecting the best of two simple algorithms, one based on the request sequence and another based on the predicted sequence. 

Further results in online algorithms with machine learned advice include the work by Lattanzi et al. \cite{LattLavaMoselVass2020}, who studied the restricted assignment scheduling problem, the work of Bamas et al. \cite{BamaMaggRohwSvens2020}, who considered energy minimization problems, and the more general framework of online primal'dual algorithms \cite{BamaMaggSven2020}. Moreover, the following online selection problems were  studied from the viewpoint of learning-augmented algorithms in \cite{AntoGoulKleeKole2020}: (i) the classical secretary problem, (ii) online bipartite matching with vertex arrivals and (iii) the graphic matroid secretary problem. 

More recently, Almanza et al. \cite{AlmanzaCLPR21} considered OFL with predictions in the form of different sets of suggested optimal facility locations. They present a randomized online algorithm with logarithmic competitiveness against an optimal solution restricted to facilities from the suggested sets. 
Azar at al. \cite{AzarPT22} considered OFL, and other online network design problems, with a prediction of the entire demand sequence is given in advance. They used cost doubling in order to combine an online algorithm applied to the actual demand sequence and an offline algorithm that computes partial solutions for the predicted input. However, their notion of prediction and error are different from ours, which makes their results and techniques incomparable to ours.  
Argue et al. \cite{AFGS22} considered OFL in a setting where the predictions are obtained by sampling an $\eps$-fraction of the demand sequence, and Gupta et al. \cite{GPSS22} considered online covering and facility location problems with predictions guaranteed to be accurate with probability at least $\eps$. 

In a different research direction, there has been significant interest recently in facility location problems with predictions from the perspective of truthful mechanism design, e.g., \cite{ABGOT22,BGTC2024,IB2022}.

%Adopting a slightly different model, Mahdian et al. \cite{MahdNazeSabe2012} studied problems where it is assumed that there exists an optimistic algorithm (which could in some way be interpreted as a prediction), and designed a meta-algorithm that interpolates between a worst-case algorithm and the optimistic one. They considered several problems, including the allocation of online advertisement space, and for each gave an algorithm whose competitive ratio is also an interpolation between the competitive ratios of its corresponding optimistic and worst-case algorithms. However, the performance guarantee is not given as a function of the ``prediction" error, but rather only as a function of the respective ratios and the interpolation parameter. In \cite{AntoGoulKleeKole2020}, the following online selection problems are studied: (i) the classical secretary problem, (ii) online bipartite matching with vertex arrivals and (iii) the graphic matroid secretary problem. 

\smallskip\noindent\textbf{Online Facility Location.} 
The (metric uncapacitated) Facility Location is a classical optimization problem
that has been widely studied in both Operations Research and Computer
Science (see e.g., \cite{DH04,Shm00}). Its online version has received significant attention since its
introduction by Meyerson \cite{Meye2001}. Fotakis \cite{Fota2008} established a lower bound of $\Omega(\log n / \log \log n)$ and showed that the competitive ratio of Meyerson's algorithm is asymptotically optimal. 
Alternative algorithms for OFL were also given using different techniques: Fotakis \cite{Fota2007} gave a deterministic primal-dual $O(\log n)$-competitive algorithm and Anagnostopoulos et al.\cite{AnagBentUpfaHent2004} gave a deterministic $O(\log n)$-competitive algorithm using a hierarchical partitioning of the metric space. 
For follow up work on OFL and its variants, we refer the reader to the survey \cite{Fota2011}. Recently, there has been research interest in the dynamic variant of OFL
\cite{CygaCzumMuchSank2018,CoheHjulParoSaulSchw2019,GuoKulkLiXian2020}.
%\cite{CoheHjulParoSaulSchw2019,GuoKulkLiXian2020}

\section{Model and Preliminaries}
\label{sec:prelim}

\noindent{\bf Notation.} 
We consider a metric space $(\mathcal{M}, d)$, where $d : \mathcal{M} \times \mathcal{M} \to \reals_{\geq 0}$ is a distance function, which is non-negative, symmetric and satisfies the triangle inequality. For a point $v \in M$ and a subset $U \subseteq M$, we let $d(v,U) = \min_{u \in U} d(v,u)$. We use the convention that $d(v, \emptyset) = \infty$.

\smallskip\noindent{\bf Online Facility Location.} 
In the Online Facility Location problem (OFL), the input consists of a demand sequence $(v_1, \ldots, v_n)$ in a metric space $(\mathcal{M}, d)$.
The demands arrive one at a time and must be assigned irrevocably to an open facility upon arrival. In response to the demand sequence $(v_1,v_2,\cdots,v_t)$, the online algorithm maintains a sequence of facility configurations $(\mathcal{F}_0, \mathcal{F}_1, \ldots, \mathcal{F}_n)$.
	
When a new demand $v_t$ arrives, the algorithm decides whether to assign it to an existing facility or to open a new one. If the algorithm opens a new facility at location $c$, the facility cost increases by $f$ and $\mathcal{F}_t = \mathcal{F}_{t-1} \cup \{c \}$. Otherwise, $\mathcal{F}_t = \mathcal{F}_{t-1}$.  Finally, the demand $v_i$ is assigned to the nearest facility in $\mathcal{F}_t$ and the assignment cost increases by $d(\mathcal{F}_t, v_t)$. The goal is to minimize the algorithm's total assignment and facility opening cost: 
\[
   \alg \coloneqq
   f\cdot |\mathcal{F}_n| + \sum_{t=1}^n d(\mathcal{F}_t, v_t) 
\]
%
%The algorithm's cost after demand $v_t$ has been processed consists of the assignment cost $\Asg_t \coloneqq \sum_{\tau=1}^t d(v_\tau,\mathcal{F}_\tau)$, which is the sum of distances of the demands up to $v_t$ to their nearest facility at their assignment time, and the facility cost $\mathrm{Fac}_t \coloneqq f\cdot |\mathcal{F}_t|$, which is the total cost of opening the facilities in $\mathcal{F}_t$\,.

We let $\mathcal{F}^\ast$ % = \{ c_1^\ast, \ldots, c_k^\ast \}$ 
denote an optimal set of facility locations for the corresponding offline instance of Facility Location with demand set $\{ v_1, \ldots, v_n \}$ (which is fully known in advance). Then, 
$\opt \coloneqq
   f\cdot |\mathcal{F}^\ast| +  \sum_{t=1}^n d(v_t,\mathcal{F}^\ast)$ 
is the optimal cost. %for the demand set $\{ v_1, \ldots, v_n \}$. 
For each demand $v_t$, $c^\ast_t = \arg\min_{c^\ast \in \mathcal{F}^\ast} d(v_t, c^\ast)$ denotes the optimal facility where $v_t$ is assigned. We note that in the optimal solution $\mathcal{F}^\ast$, for any demand $v_t$, $d(v_t, c^\ast_t) \leq f$ (since we could improve the optimal cost by opening a facility at $v_t$, otherwise). 

\smallskip\noindent{\bf Predictions.}
We consider a learning-augmented setting, where each new demand  $v_t$ is accompanied by a \emph{prediction} $p_t$ of the optimal facility $c^\ast_t$ where $v_t$ is assigned. The \emph{prediction error} of $p_t$, denoted as $\eta(t)$ or $\eta(p_t)$, is the distance of the predicted location $p_t$ to the optimal location $c^\ast_t$, i.e., 
$\eta(t) = d(p_t, c^\ast_t)$. We use the \emph{total prediction error} $\eta_1 = \sum_{t=1}^n \eta(t)$ and the \emph{maximum prediction error} $\eta_\infty = \max_t \eta(t)$ in order to quantify the inaccuracy of the predictions provided to the algorithm. 

\def\comp{cr}

\smallskip\noindent{\bf Competitive Ratio.}
We evaluate the performance of online algorithms using the \emph{competitive ratio} \cite{BoroYavi1998}. A randomized online algorithm is $\comp$-competitive if for any demand sequence (or any sequence of demand-prediction pairs), the algorithm's expected cost is at most $\comp$ times the optimal cost for the corresponding offline Facility Location instance, where the demands are fully known in advance. 

We note that the competitive ratio of an OFL algorithm with predictions may depend on the number of demands $n$, the total prediction error $\eta_1$ and the maximum prediction error $\eta_\infty$. As usual in the literature of online algorithms with predictions \cite{MitzVassSurv2020}, we pay special attention to \emph{consistency}, which is the competitive ratio when the predictions are perfect and $\eta_1 = \eta_\infty = 0$, and \emph{robustness}, which is the worst-case competitive ratio over all possible values of $\eta_1$ and $\eta_\infty$.

\smallskip\noindent{\bf Notational Conventions.} 
Demands are typically denoted by $v_t$ (or $v$), predictions by $p_t$ (or $p$). %and facilities by $c$\,. 
We sometimes refer to \emph{demand-prediction pairs} $(v_t, p_t)$, which is the input to our online algorithm, as \emph{requests}, for brevity. 
We use the term \emph{optimal center} (or \emph{center}) to refer to an optimal facility location in $\mathcal{F}^\ast$ and the term \emph{facility} to refer to an algorithm's facility in $\mathcal{F}$.
We say that a demand $v_t$ (or the pair $(v_t, p_t)$) is mapped to the center $c^\ast_t$ where $v_t$ is assigned in the optimal solution. 

%Subscripts indicate the timestep being referenced. 

When the timestep is not important or clear from the context, we omit the subscript ${}_t$ and simply use $v$ for demands and $p$ for predictions. Moreover, we usually omit the subscript in $\mathcal{F}_{t-1}$ and use $\mathcal{F}$ to denote the current set of algorithm's facilities when a new demand $v_t$ arrives. We let $\cost_{(v, p)}$ denote the algorithm's cost associated with the pair $(v, p)$, and let $\Asg^\ast_v$ (resp. $\Asg^\ast$) denote $v$'s (resp. the total) optimal assignment cost. In general, we use ${}^\ast$ to indicate the costs and facilities in the optimal solution.

%E.g., as a superscript to distinguish any quantity associated with the optimal offline solution. Again, $\Asg_{v}^{\ast}$ is the assignment cost of $v$ in the optimal offline solution. 

\section{PredOFL: Learning-Augmented Facility Location}
\label{sec:algo}

In this section, we present the $\predfl$ algorithm for OFL with predictions and establish its competitive ratio. \nameref{alg:greedy3_fl} works similarly to Meyerson's algorithm \cite{Meye2001}, but with the predictions in place of the demands. Specifically, every time a demand-prediction pair $(v_t, p_t)$ arrives, we open a facility at the predicted location $p_t$ with probability $d(p_t, \mathcal{F})/f$ (instead of $v_t$ and $d(v_t, \mathcal{F})/f$ in Meyerson's algorithm). Namely, we open a new facility with probability equal to the distance of the predicted location $p_t$ to the current set of algorithm's facilities $\mathcal{F}$ divided by the cost $f$ of opening a new facility. 
If $d(p_t, \mathcal{F}) \geq f$, we open a new facility at $p_t$ with certainty. 
The remainder of this section is devoted to upper bounding the expected cost and the competitive ratio of \nameref{alg:greedy3_fl}. %In particular, we next show that:

\begin{algorithm}[t]
\algotitle{\predfl}{alg:greedy3_fl.title}
\DontPrintSemicolon
\KwIn{\text{Sequence of demand-prediction pairs $(v_1, p_1), \ldots, (v_n, p_n)$} }
$\mathcal{F} = \emptyset$;
\tcp*{set of open facilities}
% (set of open facilities)
\ForEach{\text{demand-prediction pair $(v_t, p_t)$}}{
	%$c_{open} = \argmin\limits_{c \in \mathcal{F}} d(v, c)$\\
%    \uIf{$d(v_t,p_t)>f$}{
%		$\mathcal{F} = \mathcal{F} \cup \{ v_t \}$;
   % (new facility opens at $v_t$)
%        \tcp*{new facility opens at $v_t$}
%        \label{algortihm:if_statement}
%	}\Else{
	% $r_v = d(c_{open}, \hat{c}_v)$\\
		With probability $\min \{1, \frac{d(\mathcal{F}, p_t)}{f} \}$: \\
		\Indp $\mathcal{F} = \mathcal{F} \cup \{ p_t\}$;
        % (new facility opens at $p_t$)
        \tcp*{new facility opens at $p_t$}
        \label{algortihm:opening_prob}
	}
	Assign $v_t$ to the nearest facility in $\mathcal{F}$ with cost $d(\mathcal{F}, v_t)$;
% }
\caption{$\predfl$: Online Facility Location with Predictions}
\label{alg:greedy3_fl}	
\end{algorithm}

\subsection{Main Properties}
\label{sec:properties}

We first prove two main properties of \nameref{alg:greedy3_fl}, which are repeatedly used in the analysis of its competitive ratio. 

\begin{lemma}\label{lem:dem_cost}
Let $(v, p)$ be a demand-prediction pair mapped to optimal center $c^\ast$, and let $\fcl$ be the set of algorithm's facilities when $(v, p)$ arrives. Then, the algorithm's cost for $(v, p)$ is
\begin{align}
\E{}{\cost_{(v, p)}} & \leq \min\!\big\{ d(\mathcal{F}, p), f \big\} + d(\mathcal{F}, v)\\
\cost_{(v, p)}  & \leq f + \Asg^\ast_v + \eta(p), \hspace*{2cm} \text{if $d(\mathcal{F}, p) \geq f$} 
\end{align}
\end{lemma}

\begin{proof}
\nameref{alg:greedy3_fl} opens a facility at $p$ with probability $\min\{1, d(\fcl, p) / f \}$. If $d(\fcl, p) \geq f$, a new facility at $p$ opens with certainty and $v$'s assignment cost is $d(\fcl  \cup \{ p \}, v) \leq d(\fcl, v)$. For the latter bound, we observe that
\[ d(\fcl  \cup \{ p \}, v) \leq d(v, p) \leq d(v, c^\ast) + d(c^\ast, p) = \Asg^\ast_v + \eta(p)
\]

If $d(\fcl, p) < f$, the expected cost of \nameref{alg:greedy3_fl} for the demand-prediction pair $(v, p)$ is:

\begin{align}
\E{}{\cost_{(v, p)}}  &\leq
(f + d(\mathcal{F} \cup \{ p \}, v)) \frac{d(\mathcal{F}, p)}{f} 
+ d(\mathcal{F}, v)\left(1 - \frac{d(\mathcal{F}, p)}{f}\right) \notag \\
& \leq  d(\mathcal{F}, p) + d(\mathcal{F}, v)\,, %\label{eq:exp_cost}
\end{align}
where the last inequality holds because $d(\mathcal{F} \cup \{ p \}, v) \leq d(\mathcal{F}, v)$. 
\qed\end{proof}

\begin{lemma}[Stopping Time Lemma] \label{lem:stopping_time}
Let $\mathcal{P} = (p_1, \ldots, p_t, \ldots)$ be a sequence of predictions, where each $p_t$ causes a new facility to open at $p_t$ with probability $d(\fcl_{t-1}, p_t)/f < 1$. Then, the expected value of $\sum_{\tau=1}^{t}  d(\fcl_{\tau-1}, p_\tau)$ just before $p_{t+1}$ causes the first facility at a prediction in $\mathcal{P}$ to open is at most $f$.
\end{lemma}

\begin{proof}
For brevity, we let $d_t = d(\fcl_{t-1}, p_t)$ and $D_t = \sum_{\tau=1}^t  d(\fcl_{\tau-1}, p_\tau) = \sum_{\tau=1}^t d_\tau$ (with $D_0 = 0$) throughout the proof. We define a positive integer-valued random variable $X$ denoting the value of $D_t$ just before the point $t+1$ where $p_{t+1}$ causes the first facility at a prediction in $\mathcal{P}$ to open. Hence, $X$ takes the following values with the corresponding probability: 
\[ \begin{array}{ll}
  D_0 \ \ & \mbox{with probability\ } \tfrac{d_1}{f} \\
  D_1 & \mbox{with probability\ } 
  (1-\tfrac{d_1}{f})\tfrac{d_2}{f} \\
  D_2\ \ & \mbox{with probability\ } 
  (1-\tfrac{d_1}{f})(1-\tfrac{d_2}{f})\tfrac{d_3}{f} \\
  D_3\ \ \ \ & \mbox{with probability\ } 
   (1-\tfrac{d_1}{f})(1-\tfrac{d_2}{f})(1-\tfrac{d_3}{f})\tfrac{d_4}{f} \\
 \cdots \cdots & \cdots \cdots \\
 D_t & 
 \mbox{with probability\ } 
   \prod_{\tau=1}^t(1-\tfrac{d_\tau}{f})\tfrac{d_{t+1}}{f} \\
  \cdots \cdots & \cdots \cdots \\  
\end{array}\]

We need to show that $\E{}{X} \leq f$. We observe that for all $t \geq 0$, 
\begin{equation}\label{eq:exp-approx}
 \mathbb{P}(X > D_t) = \prod_{\tau=0}^{t+1} \left(1 - \frac{d_\tau}{f}\right)
 \leq \prod_{\tau=0}^{t+1} \exp\!\left(-\frac{d_\tau}{f}\right)
 = \exp\!\left(-\frac{D_{t+1}}{f}\right)
\end{equation} 

Therefore, by the definition of expectation for positive integer-valued random variables,
\begin{align}
 \E{}{X} & = \sum_{n=0}^\infty \mathbb{P}(X > n) 
         = \sum_{t=0}^\infty \sum_{\tau = D_t}^{D_{t+1}-1} \mathbb{P}(X > D_t) \\
        & \leq \sum_{t=0}^\infty (D_{t+1}-D_t)\,\exp(-D_{t+1}/f) \\
        & \leq \sum_{t=0}^\infty \int_{D_t}^{D_{t+1}} \exp(-x/f) dx 
           = \int_{D_0}^{\infty} \exp(-x/f) dx = f
\end{align} 
The second equality follows from the definition of $X$, because for all $t \geq 0$, $\mathbb{P}(X > n) = \mathbb{P}(X > D_t)$ for all $n \in \{ D_t, \ldots, D_{t+1}-1\}$. The first inequality follows from \eqref{eq:exp-approx}. The second inequality follows from the fact that 
$\exp(-D_{t+1}/f) \leq \exp(-x/f)$ for all $x \in [D_t, D_{t+1}]$. 
\qed\end{proof}

\subsection{Consistency: Competitive Ratio with Good Predictions}
\label{sec:consistency}

We next give an upper bound on the competitive ratio of \nameref{alg:greedy3_fl} in terms of $\frac{n \eta_\infty}{\opt}$, which is useful when predictions are very accurate.  

\begin{theorem} \label{thm:error0}
For all $\eta_\infty \geq 0$, \nameref{alg:greedy3_fl}'s competitive ratio  is at most $3\frac{n \eta_{\infty}}{\opt}+2$. 
\end{theorem}

\begin{proof}
\nameref{alg:greedy3_fl} opens a facility at $p_1$ with certainty. Since $d(c^\ast_{1}, p_1) \leq \eta_\infty$, by the definition of $\eta_\infty$, and using Lemma~\ref{lem:dem_cost}, we obtain that the algorithm's cost for $(v_1, p_1)$ is at most 
\begin{equation}\label{eq:first_demand}
   \cost_{(v_1, p_1)} \leq 
   f + d(v_1, c^\ast_1) + d(c^\ast_1, p_1)  \leq 
   f + \Asg^{\ast}_{v_1} + \eta_\infty
\end{equation}

By Lemma~\ref{lem:dem_cost}, \nameref{alg:greedy3_fl}'s expected cost for each subsequent demand-prediction pair $(v_t, p_t)$ is:
\begin{equation}\label{eq:exp_cost}
\E{}{\cost_{(v_t, p_t)}} \leq d(\mathcal{F}_{t-1},p_t) + d(\mathcal{F}_{t-1},v_t) 
\end{equation}

If when $(v_t, p_t)$ arrives, $d(\mathcal{F}_{t-1},c^\ast_t) \leq \eta_\infty$ (i.e., there is an algorithm's facility within distance $\eta_\infty$ to $c^\ast_t$), then (i) 
$d(\mathcal{F}_{t-1},p_t) \leq d(\mathcal{F}_{t-1}, c^\ast_t) + d(c^\ast_t, p_t) \leq 2\eta_\infty$; and (ii) 
$d(\mathcal{F}_{t-1},v_t) \leq d(v_t, c^\ast_t) + (\mathcal{F}_{t-1}, c^\ast_t) \leq \Asg^{*}_{v_t}
+ \eta_{\infty}$. Therefore, by \eqref{eq:exp_cost}, 
$\E{}{\cost_{(v_t, p_t)}} \leq \Asg^{*}_{v_t}+
3 \eta_{\infty}$. 

Otherwise, for each optimal center $c^\ast$, due to the facility opening rule in \nameref{alg:greedy3_fl} and Lemma~\ref{lem:stopping_time}, the expected value of $\sum_{t} d(\mathcal{F}_{t-1}, p_t)$ over all demand-prediction pairs $(v_t, p_t)$ that are mapped to $c^\ast$ and are assigned to an algorithm's facility before a facility within $\eta_\infty$ to $c^\ast$ opens is at most $f$. Therefore, using \eqref{eq:exp_cost} and that 
\begin{equation}\label{eq:close_const}
 d(\mathcal{F}_{t-1},v_t) \leq d(\mathcal{F}_{t-1},p_t) + d(p_t,c^\ast) + d(c^\ast, v_t) \leq d(\mathcal{F}_{t-1},p_t) + \eta_\infty + \Asg^\ast_{v_t}\,,
\end{equation}
we obtain that for each optimal center $c^\ast$ the expected assignment cost of all demand-prediction pairs $(v_t, p_t)$ that are mapped to $c^\ast$ and (the demands) are assigned to an algorithm's facility as along as $d(\mathcal{F}_{t-1},c^\ast) > \eta_\infty$ is at most $f$ plus their assignment cost in the optimal solution plus their number times $\eta_\infty$. Moreover, the algorithm's total cost for the first pair $(v_t, p_t)$ which is mapped to $c^\ast$ and opens a facility at $p_t$ (and thus, causes $d(\mathcal{F}_{t},c^\ast) \leq \eta_\infty$ for the first time) is bounded as in \eqref{eq:first_demand}. 

Putting everything together, we obtain that the expected total cost of \nameref{alg:greedy3_fl} is at most 
$2kf + \Asg^\ast + 3n\eta_\infty$, which divided by $\opt = kf + \Asg^\ast$ gives the desired competitive ratio. 
\qed\end{proof}

\cref{thm:error0} directly implies an upper bound of $2$ on the consistency of \nameref{alg:greedy3_fl}. 

\begin{corollary}[Consistency]
\label{cor:consistency}
If $\eta_1 = \eta_\infty = 0$ (i.e., all the predictions are perfect), the competitive ratio of \nameref{alg:greedy3_fl} is at most $2$.
\end{corollary}

\subsection{Robustness: Competitive Ratio with Arbitrary Predictions}
\label{sec:robustness}

The most interesting part of the analysis is to upper bound the competitive ratio of \nameref{alg:greedy3_fl} in case where $\opt \geq \eta_1 \geq \eta_{\infty} > \opt/n$ (i.e., the predictions are useful but may be far from perfect).

\begin{theorem}\label{thm:main}
For all sequences of demand-prediction pairs of length $n$ with $\opt \geq \eta_1 \geq \eta_\infty > \opt/n$, the competitive ratio of \nameref{alg:greedy3_fl} is at most
\begin{equation}
O\!\lp( 
\frac{\log\lp( \frac{n \eta_\infty}{\opt} \rp) }
{
\log \lp(\frac{\opt}{\eta_1} \log\lp( \frac{n \eta_\infty }{\opt} \rp) \rp)
} 
\rp)
\end{equation}
\end{theorem}

The rest of this subsection is devoted to the proof of \cref{thm:main}. For the proof, we bound the algorithm's expected total cost for each optimal cluster separately and utilize the concept of phases, a standard technique in the analysis of OFL algorithms, see e.g., \cite{Fota2011}. The approach involves defining a family of concentric balls with geometrically decreasing radii centered around each optimal center $c^\ast$. Specifically, we fix a pair of integers $m, \ell \geq 2$ so that 
\begin{equation}\label{eq:last_zone}
m^\ell \geq \frac{ n \eta_\infty}{\opt}\,,
\end{equation}
and define a family of balls $B_i(c^\ast) = \Ball(c^\ast, \frac{\eta_\infty}{ m^{i}})$ around each optimal center $c^\ast$, for all $i = 0, \ldots, \ell$. We say that a demand-prediction pair $(v, p)$ mapped to $c^\ast$ belongs to a ball $B_i(c^\ast)$, if the prediction $p \in B_i(c^\ast)$. We note that the radius of each $B_\ell$ is at most $\opt / n$, and due to the definition of $m$ and $\ell$ in \eqref{eq:last_zone}, $B_\ell(c^\ast)$ includes all demand-prediction pairs $(v, p)$ that are mapped to $c^\ast$ and have predictions $p$ with $d(c^\ast, p) \leq \opt/n$. On the other hand, the radius of each $B_0$ is $\eta_\infty$ and $B_0(c^\ast)$ includes all demand-prediction pairs $(v, p)$ mapped to $c^\ast$. 

For the analysis of \nameref{alg:greedy3_fl}, we quantify the progress of the algorithm's facilities $\fcl$ towards converging to an optimal center $c^\ast$ through phases defined using the family of balls $B_i(c^\ast)$. 

\begin{definition}[Phases]\label{def:phase}
An optimal center $c^\ast$ is in \emph{phase $i$}, for $i = 0, \ldots, \ell-1$, during the execution of \nameref{alg:greedy3_fl} as long as 
\[ \frac{\eta_\infty}{m^{i+1}} < d(\fcl, c^\ast) \leq \frac{\eta_\infty}{m^{i}} \,,
\]
is in \emph{phase $-1$} as along as $d(\fcl, c^\ast) > \eta_\infty$\,, 
and is in \emph{phase $\ell$} as along as 
\[ d(\fcl, c^\ast) \leq \frac{\eta_\infty}{m^{\ell}} \leq \frac{\opt}{n} \,. \]
\end{definition}

We partition the demand-prediction pairs that are mapped to an optimal center $c^\ast$ and arrive in phase $i$ of $c^\ast$ into \emph{close pairs} (or \emph{close requests}) and \emph{far pairs} (or \emph{far requests}) based on the distance of the predictions to $c^\ast$. 

\begin{definition}[Far and Close Requests]\label{def:far_close_demands}
A demand-prediction pair $(v, p)$ that is mapped to an optimal center $c^\ast$ and arrives in phase $i$ of $c^\ast$ is a \emph{close pair} (or a \emph{close request}), if $d(c^\ast, p) < \eta_\infty / m^{i+1}$, and a \emph{far pair} (or a \emph{far request}) otherwise. All demand-prediction pairs arriving in phase $-1$ (resp. $\ell$) of $c^\ast$ are \emph{close} (resp. \emph{far}) \emph{pairs}. 
\end{definition}

\begin{figure}[bt]
    \centering
    \includegraphics[width=0.62\textwidth]{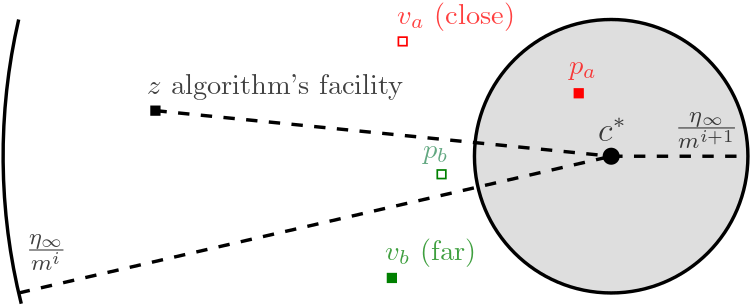}
    \caption{Phase $i$ begins when \nameref{alg:greedy3_fl} opens a facility at distance at most $\eta_\infty / m^i$ to the optimal center $c^\ast$. Demand-prediction pairs arriving in phase $i$ with predictions inside $\Ball(c^\ast, \eta_\infty / m^{i+1})$ are \emph{close pairs} (or \emph{close requests}), while the remaining demand-prediction pairs are \emph{far pairs} (or \emph{far requests}). Phase $i$ ends as soon as a close demand-prediction pair causes a new facility to open.}
    \label{fig:enter-label}
\end{figure}

We next focus on the demand-prediction pairs mapped to an optimal center $c^\ast$. We bound the expected cost of \nameref{alg:greedy3_fl} separately for the far and the close demand-prediction pairs. Lemma~\ref{lem:cost}, which is the key technical lemma of \nameref{alg:greedy3_fl}'s analysis, exploits the definition of phases in order to bound the expected cost of far and close demand-prediction pairs that are mapped to $c^\ast$ and arrive in each phase $i$, $i=-1, 0,\ldots, \ell$. 

In the following, for an optimal center $c^\ast$, we let $\costf(c^\ast)$ and $\costc(c^\ast)$ (respectively, $\asg^\ast_{far}(c^\ast)$ and $\asg^\ast_{close}(c^\ast)$, and  $\eta_{1,far}(c^\ast)$ and $\eta_{1,close}(c^\ast)$) denote the algorithm's expected cost (respectively, optimal assignment cost and the total prediction error) for far and close demand-prediction pairs mapped to $c^\ast$, respectively. Moreover, we let $\mathrm{Far}(c^\ast)$ and $\mathrm{Close}(c^\ast)$ denote the sets of far and close demand-prediction pairs mapped to $c^\ast$, respectively.

\begin{lemma}[Cost of Far and Close Requests]\label{lem:cost}
For any pair of integers $m, \ell$ that satisfy \eqref{eq:last_zone} and any optimal center $c^\ast$, the expected total cost incurred by \nameref{alg:greedy3_fl} for all far and close demand-prediction pairs mapped to $c^\ast$ is 
\begin{align}
\E{}{\costf(c^\ast)}  & \leq  \asg^*_{far}(c^\ast) + 2\,|\mathrm{Far}(c^\ast)|\tfrac{\opt}{n} + (2m+1) \eta_{1,far}(c^\ast) \label{eq:cost_far} \\
\E{}{\costc(c^\ast)} & \leq \asg^*_{close}(c^\ast) + \eta_{1,close}(c^\ast) + 2(\ell+1)f \label{eq:cost_close}
\end{align}
\end{lemma}

In the proof of Lemma~\ref{lem:cost}, the expected cost for far demand-prediction pairs is bounded by Lemma~\ref{lem:dem_cost} and the fact that their prediction error is within a factor of $m$ from their distance to the nearest algorithm's facility. As for the expected cost of close demand-prediction pairs, we again use Lemma~\ref{lem:dem_cost} and bound their cost for each phase separately. The key idea is that due to Lemma~\ref{lem:stopping_time}, their expected cost up to the point where the first facility due to a close prediction opens, which causes a new phase to begin, is at most $f$ plus additional terms due to the error of the predictions and the optimal assignment cost of the demands (see also the analysis in \eqref{eq:close_const} and the paragraph below that in Theorem~\ref{thm:error0}). We next provide the formal proof of Lemma~\ref{lem:cost}. 

\begin{proof}[of Lemma~\ref{lem:cost}]
Throughout the proof, we fix a pair of integers $m, \ell$ that satisfy \eqref{eq:last_zone} and an optimal center $c^\ast$. Our analysis is restricted to demand-prediction pairs mapped to $c^\ast$. 

\smallskip\noindent\textbf{Far Requests -- Proof of \eqref{eq:cost_far}.} 
We start with bounding the expected cost of far demand-prediction pairs mapped to $c^\ast$. Let $i \in \{0, 1, \ldots, \ell\}$ be the current phase of $c^\ast$. Then, $d(\fcl, c^\ast) \leq \frac{\eta_\infty}{m^i}$. Let $(v, p)$ be a far demand-prediction pair arriving in phase $i$ of $c^\ast$. Then, if $i < \ell$, $d(p, c^\ast) \geq \frac{\eta_\infty}{m^{i+1}}$ and 
\begin{equation}\label{eq:eta_far}
 d(\fcl, c^\ast) \leq \frac{\eta_\infty}{m^i} \leq m\,d(p, c^\ast) = m\,\eta_1(p)
\end{equation}
Otherwise, if $i = \ell$, we have that $d(\fcl, c^\ast) \leq \opt / n$. 

By Lemma~\ref{lem:cost}, we have that the algorithm's expected cost due to a far demand-prediction pair $(v, p)$ that is mapped to $c^\ast$ and arrives in phase $i < \ell$ is:
\begin{align}
  \E{}{\cost_{(v, p)}} & \leq d(\fcl, p) + d(\fcl, v)  \\
  & \leq \big(d(\fcl, c^\ast) + d(c^\ast, p)\big) + \big(d(\fcl, c^\ast) + d(c^\ast, v)\big) \\
 & = 2 d(\fcl, c^\ast) + \eta_1(p) + \asg^\ast_v \\
 & \leq (2m+1)\eta_1(p) + \asg^\ast_v \,,
\end{align}
where the last inequality follows from \eqref{eq:eta_far} and uses that $(v, p)$ arrives in phase $i < \ell$ of $c^\ast$. If $(v, p)$ arrives in phase $\ell$ of $c^\ast$, we use that $d(\fcl, c^\ast) \leq \opt / n$, instead of \eqref{eq:eta_far}, and obtain that:
\[ \E{}{\cost_{(v, p)}} \leq 2\,\opt / n + \eta_1(p) + \asg^\ast_v \]

Summing over all far demand-prediction pairs arriving in each phase $i$ of $c^\ast$ and over all phases $i = 0, \ldots, \ell$, we obtain \eqref{eq:cost_far}. 

\smallskip\noindent\textbf{Close Requests -- Proof of \eqref{eq:cost_close}.} 
We next consider the expected cost of close demand-prediction pairs mapped to the optimal center $c^\ast$. Let $i \in \{ -1, 0, \ldots, \ell-1\}$ be the current phase of $c^\ast$. Then, $\frac{\eta_\infty}{m^{i+1}} < d(\fcl, c^\ast) \leq \frac{\eta_\infty}{m^i}$. Let $(v, p)$ be a close demand-prediction pair arriving in phase $i$ of $c^\ast$. Then, $d(p, c^\ast) < \frac{\eta_\infty}{m^{i+1}}$. 

By Lemma~\ref{lem:cost}, we have that the algorithm's cost due to a close demand-prediction pair $(v, p)$ that is mapped to $c^\ast$ and is considered in phase $i = -1, 0, \ldots, \ell-1$ before the first facility due to such a demand-prediction pair opens: 
\begin{align}
  \cost_{(v, p)} & = d(\fcl, v)  
  \leq d(\fcl, p) + d(p, c^\ast) + d(c^\ast, v) 
 = d(\fcl, p) + \eta_1(p) + \asg^\ast_v
\end{align}

We note that in the upper bound on $\cost_{(v, p)}$ above, we have omitted the first term $d(\fcl, p)$ from the bound of Lemma~\ref{lem:cost}, because we care about the algorithm's cost of close demand-prediction pairs $(v, p)$ that arrive in phase $i$ and (the demands) are assigned to an algorithm's facility before a new facility due to one of them opens. So, we condition on $(v, p)$ not opening a new facility and upper bound $\cost_{(v, p)} = d(\fcl, v)$, instead of $\E{}{\cost_{(v, p)}}$. 

Due to Lemma~\ref{lem:stopping_time}, the expected value of $\sum_{(v, p)} d(\fcl, p)$ for close demand-prediction pairs $(v, p)$ that are mapped to $c^\ast$ and arrive in phase $i$ and are assigned to an algorithm's facility before a new facility due to one of them opens is at most $f$. 

The cost of the first close demand-prediction pair $(v, p)$ that is mapped to $c^\ast$, arrives in phase $i$ and causes a new facility to open at $p$ is 
\[  f + d(\fcl \cup \{ p \}, v) \leq f + d(p, c^\ast) + d(c^\ast, v) 
                                  =  f + \eta(p) + \asg^\ast_v \,. \] 
Since $(v, p)$ is a close pair and has $d(p, c^\ast) < \frac{\eta_\infty}{m^{i+1}}$, when the first such demand-prediction pair opens a new facility at $p$, the current phase $i$ ends and a new phase $i' < i$ begins. 

Summing over all close demand-prediction pairs arriving in each phase $i$ of $c^\ast$ and over all phases $i = -1, 0, \ldots, \ell-1$ with close pairs, we obtain \eqref{eq:cost_close}. 
\qed\end{proof}

Putting far and close demand-prediction pairs together, summing up over all $k$ centers in the optimal solution and using that the total number of requests is $n$, we obtain the following corollary of Lemma~\ref{lem:cost}, which bounds the total expected cost of \nameref{alg:greedy3_fl}. 

\begin{corollary}[Expected Cost of \nameref{alg:greedy3_fl}]\label{cor:cost}
For any pair of integers $m, \ell$ that satisfy \eqref{eq:last_zone}, the expected total cost incurred by \nameref{alg:greedy3_fl} is at most 
\begin{equation}\label{eq:cost_total}
2 (\ell + 1) k f + \asg^\ast + 2\,\opt + (2m+1)\eta_1 
\end{equation}
\end{corollary}

Since $\opt = k f + \asg^\ast$, the \nameref{alg:greedy3_fl}'s competitive ratio is $O(\ell + m \eta_1 / \opt)$. 

To optimize the upper bound on the competitive ratio and establish Theorem~\ref{thm:main}, we next focus on the case where $\eta_\infty > \opt / n$ (if $\eta_\infty \leq \opt / n$, the competitive ratio of \nameref{alg:greedy3_fl} is at most $5$, by Theorem~\ref{thm:error0}) and $\eta_1 \leq \opt$ (if $\eta_1 > \opt$, we should also use the demand points as possible algorithm's facility locations, see the paragraph that concludes the proof of Theorem~\ref{thm:main_intro} in Section~\ref{sec:combination}). 
%
%To optimize the upper bound on the competitive ratio and establish Theorem~\ref{thm:main}, 
We let $\ell = m \eta_1 / \opt$ and  select $m$ so that \eqref{eq:last_zone} is satisfied. Then, \eqref{eq:last_zone} becomes 
\begin{equation}\label{eq:comp_ratio}
    m^{\frac{m\eta_1}{\opt}} \geq \frac{n \eta_\infty}{\opt} \Rightarrow
  m \log m \geq \frac{\opt}{\eta_1}\log\!\lp(\frac{n \eta_\infty}{\opt} \rp).    
\end{equation}

We note that \eqref{eq:comp_ratio} gives $m = \exp(W_0(B))$, where $B=\frac{\opt}{\eta_1}\log(\frac{n \eta_\infty}{\opt})$ and $W_0(x)$ is the Lambert $W$ function. Using the bound on $W_0(x)$ in \cite{HoorHass2008}, for $x>e$ (which imposes an upper bound on $\eta_1$ with respect to $\opt$ and a lower bound on $\eta_\infty$ with respect to $\opt/ n$, so that $\frac{\opt}{\eta_1}\log(\frac{n \eta_\infty}{\opt}) > e$), we get that $m = \Theta( \frac{B}{\log B})$, which concludes the proof of Theorem~\ref{thm:main}. \qed 

\section{Combining Different OFL Algorithms - The Proof of Theorem~\ref{thm:main_intro}}
\label{sec:combination}

%$\minalg (A_0, A_1)$

To conclude the proof of Theorem~\ref{thm:main_intro}, we next show how to combine two OFL algorithms, $A_0$ and $A_1$, so that we obtain an OFL algorithm $\minalg (A_0, A_1)$, which for every sequence of demand-prediction pairs achieves a total cost within a small constant factor from the total cost of the best of $A_0$ and $A_1$ on the same sequence. 

Algorithm~\ref{alg:combine} applies the binary search approach, underlying the optimal solution to the Cow Path problem \cite{KaoRT96}, to Online Facility Location (see also \cite{AntoCoesEliaSimo2020} for a different adaptation of binary search to Metrical Task Systems). 
Algorithm~\ref{alg:combine}, also referred to as $\minalg(A_0, A_1)$, has access to OFL algorithms $A_0$ and $A_1$, and receives online a sequence $(v_1, p_1), \ldots, (v_n, p_n)$ of demand-prediction pairs. $\minalg(A_0, A_1)$ simulates both $A_0$ and $A_1$ on $(v_1, p_1), \ldots, (v_n, p_n)$, proceeds in phases guided by cost doubling, and aims to \emph{follow} (i.e., to adopt the set of facilities of) the less costly of them. Every time the total cost of the algorithm currently followed exceeds the next power of $2$, a new phase begins. While in present phase, $\minalg(A_0, A_1)$ follows the less costly of the two algorithms at the beginning of the phase.
% -- until the point where its total cost exceeds the next power of $2$ for the first time. 

\begin{algorithm}[t]
\algotitle{\minalg}{alg:combine.title}
\DontPrintSemicolon
    \KwIn{\text{OFL algorithms $A_0, A_1$; sequence of demand-prediction pairs $(v_1, p_1), \ldots, (v_n, p_n)$}}
  $\ell = 0$; $i = 0$; \tcp*{phase index $\ell$, algorithm index $i$}
  $C(A_0) = C(A_1) = 0$; 
  \tcp*{$C(A_i)$ is the total cost of $A_i$ so far}
  $\fcl_0 = \fcl_1 = \emptyset$; 
  \tcp*{$\fcl_i$ is $A_i$'s current set of facilities}
  $\fcl = \emptyset$; 
  \tcp*{$\fcl$ is $\minalg (A_0, A_1)$'s set of facilities}
  \ForEach{\text{demand-prediction pair $(v_t, p_t)$}}{
    Serve pair $(v_t, p_t)$ using $A_0$ and update $\fcl_0$ and $C(A_0)$; \\  
    Serve pair $(v_t, p_t)$ using $A_1$ and update $\fcl_1$ and $C(A_1)$; \\  
    \If{$C(A_i) > 2^\ell$}{ 
       $\ell = \ell+1$;  \tcp*{proceed to the next phase}
       \If{$C(A_i) > C(A_{1-i})$}{ 
           $i = 1-i$; %$\fcl = \fcl \cup \fcl_{i}$;
           \tcp*{switch from $A_i$ to less costly $A_{1-i}$ for the next phase}}}
     $\fcl = \fcl \cup \fcl_i$; 
     \tcp*{update $\fcl$ to current $\fcl_{i}$}\label{algo:fac_update}
     Assign $v_t$ to the nearest facility in $\fcl$ with cost $d(\fcl, v_t)$; \label{algo:asg_update}}
\caption{$\minalg(A_0, A_1)$: combining two OFL algorithms $A_0$ and $A_1$}
\label{alg:combine}	
\end{algorithm}

For the formal description of Algorithm~\ref{alg:combine} and its analysis, we let $\cost(A_i)$, $i \in \{0, 1\}$, (resp. $\cost^{\min}$) denote the total cost of algorithm $A_i$ (resp. $\minalg(A_0, A_1)$) and let $\fcl_i$ (resp. $\fcl$) denote the set of $A_i$'s (resp. of $\minalg(A_0, A_1)$'s) facilities just after the current demand-prediction pair is served. We let $\cost_t(A_i)$  (resp. $\cost_t^{\min}$) denote the total cost of algorithm $A_i$ (resp. of $\minalg(A_0, A_1)$) just after the demand-prediction pair $(v_t, p_t)$ has been served. Then, $\cost_n(A_i)$  (resp. $\cost_n^{\min}$) is the total cost of algorithm $A_i$ (resp. of $\minalg(A_0, A_1)$) for the entire request sequence. We say that $\minalg(A_0, A_1)$ \emph{follows} OFL algorithm $A_i$, $i \in \{0, 1\}$, as long as $\minalg(A_0, A_1)$ ensures that $\fcl_i \subseteq \fcl$.

$\minalg(A_0, A_1)$ proceeds in phases $\ell = 0, 1, \ldots$ determined by an upper bound of $2^\ell$ on the total cost $C(A_i)$ of the algorithm $A_i$ currently followed. At the beginning of each phase $\ell$, Algorithm~\ref{alg:combine} selects the less costly of the two algorithms $A_0$ and $A_1$ to follow. Hence, at the beginning of each phase $\ell$, $i$ is set to $1$, if $C(A_1) < C(A_0)$, and $i$ is set to $0$, if $C(A_1) > C(A_0)$ (ties are broken in favor of the algorithm currently followed). Phase $\ell$ lasts as long as $C(A_i) \leq 2^\ell$. While in phase $\ell$, Algorithm~\ref{alg:combine} maintains that $\fcl_i \subseteq \fcl$ by setting $\fcl = \fcl \cup \fcl_i$ in step \ref{algo:fac_update}, just before the current demand is served. Moreover, if $\minalg(A_0, A_1)$ switches from algorithm $A_{1-i}$ to algorithm $A_i$ at the beginning of a new phase, step \ref{algo:fac_update} ensures that every new facility added to $\fcl_i$ after $\minalg(A_0, A_1)$'s last switch from $A_i$ to $A_{1-i}$ is now added to $\fcl$. Since each demand $v_t$ is assigned to the nearest facility in $\fcl$ and $\fcl_i \subseteq \fcl$, the assignment cost $d(\fcl, v_t)$ of $\minalg(A_0, A_1)$ for each demand $v_t$ is at most the assignment cost $d(\fcl_i, v_t)$ of the algorithm $A_i$ currently followed.

The competitive ratio of $\minalg(A_0, A_1)$ follows from the fact that at any point in time, the total cost of $\minalg(A_0, A_1)$ is at most the total cost of the algorithm $A_i$ currently followed plus the total cost of the algorithm $A_{1-i}$ up to the last demand-prediction pair where $\minalg(A_0, A_1)$ followed $A_{1-i}$. This is summarized in Proposition~\ref{prop:combine-cost}, whose proof is by induction on $t$ and is deferred to Section~\ref{sec:a:prop:combine-cost}. 

\begin{proposition}\label{prop:combine-cost}
For $t \geq 1$, let $t_l \leq t$ be the last step no later than $t$ where $\minalg(A_0, A_1)$ switches from algorithm $A_{1-i}$ to algorithm $A_i$ ($t_l$ may be $t$). Then, at the end of step $t$, $\fcl_i \subseteq \fcl$ and 
\begin{equation}\label{eq:switch}
 C_t^{\min} \leq C_t(A_i) + C_{t_l-1}(A_{1-i}) 
\end{equation}
\end{proposition}

We are now ready to prove the main result of this section:

\begin{theorem}\label{thm:combine}
Let $\algo{0}$ and $\algo{1}$ be algorithms for Online Facility Location with predictions. Then, for every sequence of $n$ demand-prediction pairs, the total cost of $\minalg (\algo{0}, \algo{1})$, formally described in Algorithm~\ref{alg:combine}, is 
\[ \cost^{\min}_n \leq 3 \min\{ \cost_n(\algo{0}), \cost_n(\algo{1}) \}\,. \]
\end{theorem}

\begin{proof}
The proof follows from Proposition~\ref{prop:combine-cost} and the fact that the most costly of the two algorithms $A_0$ and $A_1$ cannot contribute more than twice the total cost of $\min\{ \cost_n(\algo{0}), \cost_n(\algo{1}) \}$ to \eqref{eq:switch}. 
More precisely, let $A_0$ 
be the less costly algorithm on the entire sequence and let $\ell \geq 0$ be such that $2^\ell < \cost_n(\algo{0}) \leq 2^{\ell+1}$. 

If $\minalg (\algo{0}, \algo{1})$ follows $A_0$ at the end of step $n$, by Proposition~\ref{prop:combine-cost}, 
$\cost^{\min}_n \leq C_n(A_0) + C_{n_l-1}(A_1)$. We note that $C_{n_l-1}(A_1) \leq 2^{\ell+1}$, because if at some step $t$, $C_{t}(A_1) > 2^{\ell+1}$, $\minalg (\algo{0}, \algo{1})$ switches to the less costly algorithm $A_0$ and keeps following $A_0$ until step $n$. Therefore, since $C_n(A_0) > 2^{\ell}$, 
\[ \cost^{\min}_n \leq C_n(A_0) + C_{n_l-1}(A_1) 
                  \leq C_n(A_0) + 2^{\ell+1} 
                  < 3\,C_n(A_0)\,. \]

If $\minalg (\algo{0}, \algo{1})$ follows $A_1$ at the end of step $n$, by Proposition~\ref{prop:combine-cost}, 
$\cost^{\min}_n \leq C_{n_l-1}(A_0) + C_{n}(A_1)$. As before, we note that $C_{n}(A_1) \leq 2^{\ell+1}$. %, since if at some step $t$, $C_{t}(A_1) > 2^{\ell+1}$, $\minalg (\algo{0}, \algo{1})$ switches to the less costly algorithm $A_0$ and keeps following $A_0$ until the end of the sequence. 
Therefore, 
\[ \cost^{\min}_n \leq C_{n_l-1}(A_0) + C_{n}(A_1) 
                  \leq C_n(A_0) + 2^{\ell+1} 
                  < 3\,C_n(A_0)\,, \]
because $C_n(A_0) > 2^{\ell}$. 
\qed\end{proof}

For randomized algorithms $A_0$ and $A_1$, the bounds of Proposition~\ref{prop:combine-cost} and Theorem~\ref{thm:combine} hold for the algorithms' realized cost on any request sequence. Then, we get that the expected cost of $\minalg (\algo{0}, \algo{1})$ is at most $3$ times the minimum of the expected costs of $A_0$ and $A_1$.

\smallskip\noindent\textbf{The Proof of Theorem~\ref{thm:main_intro}.}
To obtain the competitive ratio of Theorem~\ref{thm:main_intro}, we apply Theorem~\ref{thm:combine} to the algorithm \nameref{alg:greedy3_fl} and to the algorithm of \cite{JiangLLTZ22} for uniform facility costs. 
The latter algorithm is complementary to \nameref{alg:greedy3_fl} in the sense that its decisions for opening new facilities are mostly guided by the demand locations. Specifically, for each demand-prediction pair $(v_t, p_t)$, the algorithm opens a pair of new facilities, one at $v_t$ and another at $p_t$, with probability $\min\{ d(\fcl, v_t) / f, 1 \}$. 

For that algorithm, we can prove that $\E{}{\cost_{(v, p)}} \leq \min\{2f, 3d(\fcl, v)\}$ for any demand-prediction pair $(v, p)$ (the proof is similar to Lemma~\ref{lem:dem_cost}). By defining \eqref{eq:last_zone} and phases wrt. $\frac{n}{\opt}\min\{ \eta_\infty, f \}$, using that $\Asg^\ast_v \leq f$ for any demand $v$, and adapting the proof of Lemma~\ref{lem:cost}, we can show that for any integers $m, \ell \geq 2$ that satisfy $m^\ell \geq \frac{n}{\opt}\min\{ \eta_\infty , f \}$, the total expected cost of the algorithm of \cite{JiangLLTZ22} for uniform facility costs is at most
\[ 3(\ell+1) k f + 3(m+1)\Asg^\ast + 3\,\opt \] 
Then, working as in the last paragraph of Section~\ref{sec:algo} and using that $\opt \geq f$, we can show that the algorithm's competitive ratio for uniform facility costs is 
\begin{equation}\label{eq:std_comp}
O\!\left(\frac{\log\!\left(\tfrac{n}{\opt}\min\{ \eta_\infty , f \}\right)}{\log\log\!\left(\tfrac{n}{\opt}\min\{ \eta_\infty , f \}\right)}\right) = 
O\!\left(\frac{\log \min\!\left\{ \tfrac{ n \eta_\infty}{\opt}, n \right\}}{\log\log \min\!\left\{ \tfrac{ n \eta_\infty}{\opt}, n \right\}}\right)\,,
\end{equation} 
which is independent of the total prediction error $\eta_1$. 

Theorem~\ref{thm:combine} shows how to get an online algorithm with thrice the minimum of the competitive ratio of \nameref{alg:greedy3_fl} (as established in Theorem~\ref{thm:error0} for small $\eta_\infty$ and in  Theorem~\ref{thm:main} for $\opt \geq \eta_1 \geq \eta_\infty > \opt / n$) and the competitive ratio in \eqref{eq:std_comp}, which is achieved by algorithm of \cite{JiangLLTZ22} for uniform facility costs and holds for all $\eta_1 \geq 0$. Hence, using Algorithm~\ref{alg:combine} and Theorem~\ref{thm:combine}, we obtain the competitive ratio of Online Facility Location with predictions stated in Theorem~\ref{thm:main_intro}. \qed

%\endinput

\subsection{The Proof of Proposition~\ref{prop:combine-cost}}
\label{sec:a:prop:combine-cost}

\begin{proof}
The proof is by induction on $t$. For the base, we consider $t=1$. If $(v_1, t_1)$ does not cause a switch, $C_1^{\min} = C_1(A_0)$ and \eqref{eq:switch} holds. If $(v_1, t_1)$ causes a switch from algorithm $A_0$ to algorithm $A_1$, $C_1^{\min} = C_1(A_1)$ and \eqref{eq:switch} holds. In both cases, $\minalg(A_0, A_1)$ concludes its first step with $\fcl_i = \fcl$. 

We inductively assume that \eqref{eq:switch} and $\fcl_i \subseteq \fcl$ hold at the end of every step $t \geq 1$. We next prove that both properties hold at the end of the next step $t+1$.

If $(v_{t+1}, t_{t+1})$ does not cause a switch, then $i$ and $t_l$ do not change from step $t$ to step $t+1$. The induction hypothesis is that $C_{t}^{\min} \leq C_{t}(A_i) + C_{t_l-1}(A_{1-i})$ and that $\fcl_i \subseteq \fcl$ at the end of step $t$. We observe that the additional cost of $\minalg(A_0, A_1)$ for the demand-prediction pair $(v_{t+1}, p_{t+1})$ is at most the additional cost of algorithm $A_i$ for $(v_{t+1}, p_{t+1})$. Specifically, in step~\ref{algo:fac_update}, since at the end of step $t$, $\fcl_i \subseteq \fcl$, a new facility is added to $\fcl$ only if the same facility is added to $\fcl_i$ due to $(v_{t+1}, p_{t+1})$. Moreover, setting $\fcl = \fcl \cup \fcl_i$ in step~\ref{algo:fac_update} ensures that $\fcl_i \subseteq \fcl$ at the end of step $t+1$. Hence, when Algorithm~\ref{alg:combine} reaches step~\ref{algo:asg_update}, we have that $d(\fcl, v_{t+1}) \leq d(\fcl_i, v_{t+1})$. 

If the demand-prediction pair $(v_{t+1}, t_{t+1})$ causes a switch, then $t_l = t+1$. For clarity and without loss of generality, we assume that $\minalg(A_0, A_1)$ switches from $A_0$ to $A_1$ during the step $t+1$ (i.e., $i = 0$ at the end of step $t$ and $i$ is set to $1$ during step $t+1$). We let $t'_l \leq t$ be the last step where $\minalg(A_0, A_1)$ switches from algorithm $A_1$ to algorithm $A_0$. Then, by the induction hypothesis, we have that (i) $C_{t}^{\min} \leq C_{t}(A_0) + C_{t'_l-1}(A_{1})$; (ii) at the end of step $t$, $\fcl_0 \subseteq \fcl$; and (iii) at the end of step $t'_l-1$, $\fcl_1 \subseteq \fcl$.

After the switch from $A_0$ to $A_1$ caused by $(v_{t+1}, t_{t+1})$, we have that $C_{t_l - 1}(A_0) = C_{t}(A_0)$, because $t_l = t+1$. The additional cost of $\minalg(A_0, A_1)$ due to the update of $\fcl = \fcl \cup \fcl_1$ in step~\ref{algo:fac_update} and the assignment of the demand $v_{t+1}$ in step~\ref{algo:asg_update} is at most $C_{t+1}(A_{1}) - C_{t'_l-1}(A_{1})$. To see this, we recall that at the end of step $t'_l-1$, we have that $\fcl_1 \subseteq \fcl$. Hence the update $\fcl = \fcl \cup \fcl_1$ just before the end of step $t+1$ adds to $\fcl$ (and increases the facility cost of $\minalg(A_0, A_1)$ by the cost of) the facilities added to $\fcl_1$ in the steps $t'_l, \ldots, t+1$. The total cost of these facilities is accounted in the facility cost of the difference $C_{t+1}(A_{1}) - C_{t'_l-1}(A_{1})$. The assignment cost of the demand $v_{t+1}$ is accounted in the assignment cost of the difference $C_{t+1}(A_{1}) - C_{t'_l-1}(A_{1})$, because at $v_{t+1}$'s assignment time $\fcl_1 \subseteq \fcl$ and $d(\fcl, v_{t+1}) \leq d(\fcl_1, v_{t+1})$. 

Putting everything together, we have that at the end of step $t+1$, $\fcl_i \subseteq \fcl$ and 
\begin{align}
 C^{\min}_{t+1} & \leq C_{t}(A_0) + C_{t'_l-1}(A_{1}) + 
                       (C_{t+1}(A_{1}) - C_{t'_l-1}(A_{1})) \\
                & = C_{t+1}(A_{1}) + C_{t_l-1}(A_0) \\
                & = C_{t+1}(A_{i}) + C_{t_l-1}(A_{1-i}) 
\end{align}
The inequality follows from the induction hypothesis that $C_{t}^{\min} \leq C_{t}(A_0) + C_{t'_l-1}(A_{1})$ and from the upper bound above on the additional cost of $\minalg(A_0, A_1)$ during step $t+1$. The first equality uses that $C_{t_l - 1}(A_0) = C_{t}(A_0)$, because $t_l = t+1$. The second inequality uses our hypothesis that $\minalg(A_0, A_1)$ switches from $A_0 = A_{1-i}$ to $A_1 = A_i$ during the step $t+1$. 
\qed\end{proof}

\section{Lower Bound}
\label{sec:lower_bound}

%After presenting a learning augmented algorithm, one might reasonably wonder whether it achieves the optimal competitive ratio. That is, if there is an alternative algorithm that  achieves even lower competitive ratio. Fortunately,  we prove that it is not possible up to constants, by providing a lower bound on any randomized algorithm that uses similar predictions.

We proceed to establish that the dependence of $\predfl$'s competitive ratio on $\eta_1$ (and also $\eta_\infty$ and $n$) is essentially optimal.
%
%The following generalizes the lower bound for OFL (without predictions) presented in \cite{Fota2008}. %and extends the lower bound for OFL with predictions presented in \cite[Theorem~F.1]{JiangLLTZ22}. 

\begin{theorem}\label{thm:lower_bound}
For every integer $n$ large enough, any $\alpha \in (27/n, 1/3)$ and any $\beta \in (3\alpha, 1]$ ($\alpha$ and $\beta$ may depend on $n$), there are Online Facility Location instances with predictions with $n$ demand-prediction pairs, $\eta_\infty / \opt = \alpha$ and $\beta/3 \leq \eta_1 / \opt \leq \beta$ where any randomized algorithm has competitive ratio at least
	\[
			\Omega\!\lp( \frac{\log\!\lp( \frac{n \eta_\infty}{\opt} \rp) }{\log\!\lp( \frac{\opt}{\eta_1} \log\!\lp(  \frac{n \eta_\infty}{\opt}\rp) \rp) } \rp) =
			\Omega\!\lp( \frac{\log(\alpha n) }{\log\!\left(\frac{\log(\alpha n)}{\beta}\right) } \rp) 
	\]
\end{theorem}

\begin{proof}
We generalize the lower bound for OFL (without predictions) presented in \cite{Fota2008}. Using Yao's principle (see e.g., \cite[Chapter 8.4]{BoroYavi1998} and \cite{Yao1977}), we obtain the lower bound by considering the expected cost of any deterministic algorithm against an appropriately constructed probability distribution on sequences of demand-prediction pairs of length $n$. 
We note that the lower bound is most interesting for $\alpha$ in $o(1)$ (e.g., for $\alpha = 1/\mathrm{poly}(\log n)$ or for $\alpha = n^{-\eps}$) and for $\beta$ not very much larger than $\alpha$. Hence, for clarity and by assuming that $n$ is appropriately large, we sometimes treat multiples of $1/\alpha$, $\alpha$, $1/\beta$ or $\beta$ as integers (ignoring the rounding error). %and also sometimes use $\approx$ (instead of using strict equality). It is not hard to verify that these approximations do not affect the validity of the asymptotics of the lower bound.

The metric space is a binary Hierarchical Well-Separated Tree $T$. Given $n$, $\alpha$ and $\beta$, we select $m \geq \max \{ 1/\beta, 4 \}$ and $\ell = \lceil \beta m \rceil$ so that $m$ is the least integer that satisfies $m^{\ell} \geq \alpha n$. $T$ has $\ell+1 \geq 2$ levels. We consider $T$'s root as being at level $0$ and $T$'s leaves as being at level $\ell$. For each vertex $v$, we let $T_v$ denote the subtree of $T$ rooted at $v$. 

The distance of the root to its children is $\frac{m^{\ell}}{\beta}$, and the edge lengths along each path from the root to a leaf decrease by a factor of $m$ at every level. Hence, the distance of any vertex of $T$ at level $i = 0, \ldots, \ell-1$ to its children is $m^{\ell-i}/\beta$. Observe that the following properties hold for any level-$i$ vertex $v_i$:
\begin{enumerate}[label={\arabic*}., ref={\arabic*}]
\label{prop:lb}
\item \label{prop:lb:prop1}
The distance of $v_i$ to any vertex in $T_{v_i}$ is at most $\frac{m^{\ell - i}}{\beta} \frac{m}{m-1}$.

\item \label{prop:lb:prop2}
The distance of $v_i$ to any vertex not in	$T_{v_i}$ is a least $m^{\ell+1 - i}/\beta$.
\end{enumerate}
The construction is depicted in \cref{fig:hst}.%, in \cref{a:sec:fig_lower_bound}. 

	\begin{figure}[ht]
		\centering
		 \begin{tikzpicture}[level 1/.style={sibling distance=5em}, level 3/.style={sibling distance=8em},level 4/.style={sibling distance=5em}, scale=0.8]
 \tikzset{My Style/.style={shape=circle, draw=black}}
 \tikzset{Tv1/.style={shape=circle, draw=black, fill=gray}}
 \tikzset{NotTv1/.style={shape=circle, draw=black}}

 \node [My Style,label={[yshift=-0.5cm,xshift=1cm]$\textcolor{red}{\leftarrow \frac{1}{\alpha}}$}] {}

 child{ node[My Style](b) {} 
 	child{ node[](b1) {} edge from parent [draw=none]}
 	edge from parent node[left] {} 
 } %root's left childf
 child{ node [My Style, label={[yshift=-0.5cm, xshift=1cm]$\textcolor{red}{\leftarrow \frac{m}{\alpha}}$}](c) {} 
 	child{node [](c1) {}
 		edge from parent [draw=none]}
 	child{node[NotTv1, label={[yshift=-0.5cm, xshift=1.1cm]$T_{v_i}$\hspace{0.2cm} $ \textcolor{red}{\leftarrow \frac{m^i}{\alpha}}$}](d) {} edge from parent [draw=none]
 		child{ node [NotTv1](e) {} 
 			child{ node [NotTv1](e1) {}
 				child{ node[](e11) {} edge from parent[draw=none]
 				}
 			}
 			child{ node [NotTv1](e2) {}
 				child{ node[](e21) {} edge from parent[draw=none]
 				}
 			}
 		node [left] {$T_{v_i}\setminus T_{v_{i+1}}$}
 		}%e
 		child{ node [Tv1, label={[yshift=-0.5cm, xshift=1.3cm]$T_{v_{i+1}}$\hspace{0.2cm} $ \textcolor{red}{\leftarrow \frac{m^{i+1}}{\alpha}}$}](f) {} 
 			child{ node [Tv1](g) {}
 				child{ node[](g1) {} edge from parent [draw=none]
 				}
 			}%g
 			child{ node [Tv1](h) {}
 				child{ node[](h1) {} edge from parent [draw=none]
 					}%h1 left invisible child
 				child{ node[Tv1](h2) {} edge from parent [draw=none]
 					child{ node[Tv1](final1) {}
 						}%level h node left
 					child{ node[Tv1, label={[yshift=-0.5cm, xshift=1.2cm]$\textcolor{red}{\leftarrow \frac{m^{\ell}}{\alpha}}$}](final2) {} 
 							edge from parent node[right, xshift=-0.2cm]{\textcolor{blue}{\textbullet $\hat{f}_{\ell-1}$}  \hspace{0.1cm}$\frac{m}{\beta}$}
 						}%level h node right
 					}%h2 right child
 					edge from parent node[right, xshift=-0.2cm]{\textcolor{blue}{\textbullet $\hat{f}_{i+1}$}  \hspace{0.1cm}$\frac{m^{\ell-(i+1)}}{\beta}$}
 				}%h
 			edge from parent node[right, xshift=-0.2cm]{\textcolor{blue}{\textbullet $\hat{f}_i$}  \hspace{0.1cm}$\frac{m^{\ell-i}}{\beta}$}
 			}
 		}%node d
 	edge from parent node[right, xshift=-0.2cm]{\textcolor{blue}{\textbullet $\hat{f}_0$}  \hspace{0.1cm}$\frac{m^{\ell}}{\beta}$}
 	}; %node c

 \path (c) -- node[auto=false]{\ldots} (d);
 \path (c) -- node[auto=false]{\ldots} (c1);
 \path (b) -- node[auto=false]{\ldots} (b1);
 \path (h) -- node[auto=false]{\ldots} (h2);
 \path (e21) -- node[auto=false]{\ldots} (e2);
 \path (e11) -- node[auto=false]{\ldots} (e1);
 \path (g1) -- node[auto=false]{\ldots} (g);

%Draw level labels
\node[] at (-3,0) {Level $0$};
\node[] at (-3,-1.5) {Level $1$};
\node[] at (-3,-3) {Level $i$};
\node[] at (-3,-4.5) {Level $i+1$};
\node[] at (-3,-6) {Level $i+2$};
\node[] at (-3,-9) {Level $\ell $};
 \end{tikzpicture}
		\caption{
		A sketch of the lower bound construction in the proof of Theorem~\ref{thm:lower_bound}. We depict (i) the demand locations (and their number) in each phase with red color; (ii) the tree layers (on the left), the distances of each vertex to its children (on the right) and the subtrees with text in black; and (iii) the locations of the predictions in each phase with blue color.}
		\label{fig:hst}
	\end{figure}

\smallskip\noindent{\bf Demand Sequence.} 
The demand sequence is divided into $\ell+1$ phases. Phase $0$ consists of $1/\alpha$ demands located at the root $v_0$ of $T$ (if necessary, we round up the number of demands to the nearest integer). After the end of phase $i-1$, $i=1,\ldots,\ell$, the adversary selects $v_{i}$ uniformly at random from the two children of $v_{i-1}$. In the next phase $i$, $\frac{m^i}{\alpha}$ demands arrive at the vertex $v_{i}$. The total number of demands is: 
\begin{equation}\label{eq:no_of_demands}
		\sum_{i=0}^{\ell} \frac{m^i}{\alpha} 
			= \frac{1}{\alpha} \frac{m^{\ell+1}-1}{m-1} 
			\geq \frac{m^{\ell}}{\alpha}
\end{equation}
By removing demands from the last phase, we ensure that the total number of demands is $n \leq m^\ell / \alpha$. 

\smallskip\noindent{\bf Facility Opening Cost.} 
We set $f = \frac{m-2}{m-1}\cdot \frac{m^{\ell+1}}{\alpha}$.

\smallskip\noindent{\bf Optimal Cost.} 
The optimal solution opens a single facility at $u_\ell$, i.e., at the leaf of $T$ where the demands of the last phase are located (and thus, there is no assignment cost for the demands in phase $\ell$). By Property~\ref{prop:lb:prop1}, in each phase $i = 0, \ldots, \ell-1$, the optimal solution incurs an assignment cost no larger than
\[ 		
  \frac{m^i}{\alpha} \, \frac{m^{\ell-i}}{\beta} \frac{m}{m-1} = \frac{m^{\ell}}{\alpha\beta} \frac{m}{m-1} \,,\]
because for each demand location $v_i$, the demands of all subsequent phases are located within $T_{v_i}$\,. %Since there is no assignment cost for the demands in phase $\ell$, and 
Using that $\ell = \beta m$, that $f = \frac{m-2}{m-1}\cdot \frac{m^{\ell+1}}{\alpha}$ and that $m \geq 4$, we obtain 
\begin{equation}\label{eq:lb_cost_opt}
    \opt \leq f + \ell\,\frac{m^{\ell}}{\alpha \beta} \frac{m}{m-1} 
           =  \frac{m^{\ell+1}}{\alpha}   \frac{m-2}{m-1} 
              + \frac{m^{\ell+1}}{\alpha} \frac{m}{m-1} 
          = 2\,\frac{m^{\ell+1}}{\alpha} 
          \leq 3\, \frac{m^{\ell+1}}{\alpha} \frac{m-2}{m-1} = 3f
%    f + \frac{m^{\ell+1}}{\alpha} \frac{m}{m-1} = f\,\frac{2m-1}{m-1}\,, 
\end{equation} 
%
%using that $\ell = m\alpha$ and that $f=\frac{m^{\ell+1}}{\alpha}$. 
%
Moreover, the optimal assignment cost $\Asg^\ast$ is at least $m^{\ell+1} / \alpha$, which implies that $\Asg^\ast \geq \opt / 2$, and at most $\frac{m^{\ell+1}}{\alpha} \frac{m}{m-1}$, which implies that $\Asg^\ast \leq 2\opt / 3$ (where we use that $m \geq 4$).  

\smallskip\noindent{\bf Prediction Sequence.} 
Similarly to the demand sequence, the prediction sequence is divided into $\ell+1$ phases. For the first demand of phase $0$ located at $v_0$, the corresponding prediction is at distance $\eta_\infty = \alpha \opt \leq 2m^{\ell+1}$ to $v_\ell$ (if $\alpha \opt > d(v_0, v_{\ell})$, we add a new root $v'_0$ to $T$ at distance $\eta_\infty$ to $v_\ell$ and let the corresponding prediction be located at $v'_0$). 

For each demand of phase $i = 0, \ldots, \ell-1$ located at $v_i$ (with the exception of the first demand of phase $0$), the corresponding prediction is at distance $\beta d(v_i, v_\ell)$ to $v_\ell$ (which is the leaf where the optimal facility is located). 
%
%Hence those predictions are $1/\beta$ times closer to the optimal facility location than the corresponding demands. 
%
The predictions for the demands of phase $\ell$ are located at $v_\ell$. Since $\beta \geq 1/m$, the predictions for the demands located at $v_i$ are located along the edge connecting $v_i$ to the location $v_{i+1}$ of the demands of phase $i+1$, as also shown in \cref{fig:hst}. 

The total prediction error is $\eta_1$ is at most $\alpha \opt$, due to the prediction accompanying the first demand of phase $0$, plus at most $\beta \Asg^\ast$, due to the predictions accompanying the remaining demands. Therefore, using that $\beta \geq 3\alpha$, we obtain 
\[ \eta_1 \leq \alpha \opt + \beta \Asg^\ast \leq (\alpha + 2\beta /3) \opt \leq \beta \opt \]

On the other hand, since $\alpha \leq 1/3$, there are at least $3$ demands in phase $0$. Therefore, the optimal assignment cost of the demands arriving after the first one is at least $2\Asg^\ast / 3 \geq \opt /3$. Therefore, the total prediction error $\eta_1 \geq \beta \opt / 3$. 

\smallskip\noindent{\bf Algorithm's Cost.} 
We let $\alg$ denote any deterministic algorithm (and by abusing the notation we let $\alg$ also denote the algorithm's cost).  
At the end of any phase $i$, $\alg$ knows that the optimal solution opens a facility in $T_{v_i}$ (wlog. we can assume that facilities open at the leaves of $T$), but $\alg$ cannot tell the particular leaf where the optimal facility is located. 

To estimate the algorithm's cost, we fix the adversary's choices up to phase $i$ and consider the assignment cost incurred by $\alg$ for the demands and the facilities not in $T_{v_{i+2}}$ (since the predictions of the demands in $v_i$ provide information about the location $v_{i+1}$ of the demands arriving in phase $i+1$, but they do not provide any information about the location of $v_{i+2}$ whatsoever). We distinguish between two cases: 

\begin{enumerate}[label=({\arabic*})]
\item  $\alg$ has no facilities in $T_{v_i}$ when the first demand at $v_{i+2}$ arrives. Then, the assignment cost for the demands at $v_{i},v_{i+1} \in T_{v_i} \setminus T_{v_{i+2}}$ is at least 
\[ \frac{m^{i}}{\alpha} \frac{m^{\ell+1-i}}{\beta} + 
   \frac{m^{i+1}}{\alpha} \frac{m^{\ell-i}}{\beta} = 
					 \frac{2\,m^{\ell+1}}{\alpha \beta} > 
					 2\,\frac{m-2}{m-1}\cdot \frac{m^{\ell+1}}{\alpha} = 
					 2f\,,\]
where the last inequality follows from $\beta < 1$. 
     
\item $\alg$ has at least one facility in $T_{v_i}$ when the first demand at $v_{i+2}$ arrives. In fact, we can assume that $\alg$ has at least one facility in the subtree $T_{v_{i+1}}$, for which information is provided to the algorithm by the predictions of the demands in phase $i$. Then, with probability $1/2$, the adversary selects $v_{i+2}$ so that at least one of $\alg$'s facilities in $T_{v_{i+1}}$ is not included in $T_{v_{i+2}}$. Therefore, $\alg$ incurs an expected facility cost of at least $f/2$ for facilities in $T_{v_i} \setminus T_{v_{i+2}}$. 
\end{enumerate}

Taking into account the first $\ell$ phases, we get that the expected algorithm's cost is at least $\ell f / 4$. Since $\opt \leq 3f$ and the expected algorithm's cost is at least $\ell f / 4$, the resulting competitive ratio is $\Omega(\ell) = \Omega(\beta m)$. We recall that $\ell = \beta m$ and that $m$ is chosen so that $m^{\beta m} \geq \alpha n$. Therefore, using that $\alpha = \eta_\infty / \opt$ and that $\beta \leq 3\eta_1 / \opt$, we obtain that
\begin{equation}\label{eq:lb_comp_ratio}
   m^{\beta m} \geq \alpha n \Rightarrow 
   m \log m \geq \frac{\log(\alpha n)}{\beta} \Rightarrow 
   m \log m \geq \frac{3 \opt}{\eta_1}\log\!\left(\frac{n \eta_\infty}{\opt}\right)
\end{equation}

We note that \eqref{eq:lb_comp_ratio}, which determines the lower bound on the competitive ratio of any randomized algorithm for OFL with predictions as a function of $n$, $\beta \approx \frac{\eta_1}{\opt} < 1$ and $\alpha n = \frac{n\eta_\infty}{\opt} < n/3$, is essentially identical to \eqref{eq:comp_ratio}, which determines the upper bound on the competitive ratio of $\predfl$. Working as in Section~\ref{sec:algo}, we obtain that for $B=\frac{3\opt}{\eta_1}\log(\frac{n \eta_\infty}{\opt}) > e$, \eqref{eq:lb_comp_ratio} is satisfied by $m = \Omega( \frac{B}{\log B})$, which implies a lower bound of $\Omega(\beta m) = \Omega(\beta \frac{B}{\log B})$. Using $\beta \geq \eta_1 / \opt$ and $B=\frac{3\opt}{\eta_1}\log(\frac{n \eta_\infty}{\opt})$ in $\Omega(\beta \frac{B}{\log B})$, we obtain the desired lower bound on the competitive ratio. 
\qed \end{proof}

\begin{remark}\label{rem:lower_bound}%[Comparison to \cite[Theorem~F.1]{JiangLLTZ22}]
The lower bound of Theorem~\ref{thm:lower_bound} can be regarded as a refined version, also parameterized by $\eta_1/\opt$, of \cite{JiangLLTZ22}'s lower bound. Specifically, in the lower bound of \cite[Theorem~F.1]{JiangLLTZ22}, which is based on a similar metric space and demand sequence, for any fixed $\alpha = \eta_\infty / \opt \in (0, 1]$, the predictions corresponding to the demands at the first levels of $T$ (those closer to the root) are located at distance $\eta_\infty$ to $v_\ell$, so that the maximum prediction error takes the desired value. For the remaining requests, \cite[Theorem~F.1]{JiangLLTZ22} places the predictions at the same locations as the corresponding demand points, which results in a total prediction error $\eta_1 = \Theta(\opt)$, even if $\alpha = \eta_\infty / \opt$ is very small, e.g., even if $\eta_\infty = \opt / n^{(1-\delta)}$, for any constant $\delta > 0$. Hence, the lower bound of \cite{JiangLLTZ22} does not quantify how fast the competitive ratio of OFLpred can improve as $\eta_1/\opt$ decreases (assuming a fixed value of $\alpha = \eta_\infty / \opt$). Thus, it fails to differentiate, as far as their best possible competitive ratio is concerned, between instances described in cases (i)-(iii) in the beginning of Section~\ref{subsec:contribution}. To close this gap, Theorem~\ref{thm:lower_bound} establishes a lower bound on the best possible competitive ratio of Online Facility Location with predictions which for every fixed $\alpha = \eta_\infty / \opt$, is also parameterized by $\beta \approx \frac{\eta_1}{\opt}$ and can be applied to tell such instances apart as far as their best possible competitive ratio is concerned. \qed
\end{remark}

\section{Experimental Evaluation}
\label{sec:experiments}

We next describe our experimental setup, the datasets used in our experimental evaluation, our prediction generation approach, and we comment on the experimental results. 

%We next offer a brief overview of our experimental setup and results ; %a detailed exposition is deferred to \cref{sec:appendix_experiments}.
 
%Our experimental evaluation was carried out using the standard CoverType~\cite{CovType} and US Census \cite{USCensus} datasets, which have been previously used \cite{CoheHjulParoSaulSchw2019}. Additionally, we constructed a synthetic dataset generated by uniformly sampling points on a grid.

\smallskip\noindent\textbf{Datasets.}
We used the following datasets, also used in \cite{CoheHjulParoSaulSchw2019}. 

\begin{itemize}
	\item  The CoverType dataset \cite{CovType}, from the UCI repository with 58K demand points in 54 dimensions. 
	
    \item The US Census	dataset \cite{USCensus}, from the UCI repository with 2.5M demand points in 68 dimensions. 

    \item  A synthetic dataset, created by sampling 2K points uniformly at random on the grid $[10^6, 10^6]$. 
\end{itemize}

All datasets are equipped with the Euclidean metric. The facility cost has been set to half the diameter of the underlying metric space. For the datasets of \cite{CovType,USCensus}, we restricted our experiments to the first 20K points. 

The code was written in Python and the experiments were executed on a Debian virtual machine in Google Cloud with 16vCPUs and 30GB of memory. The code, the datasets and some of the results can be found in \url{ https://anonymous.4open.science/r/PredOFL-62F1}\,.

\smallskip\noindent\textbf{Predictions.} 
To generate the predictions, we first compute the optimal offline solution in order to determine the locations of the optimal centers. For the optimal solution, we solve the Facility Location LP-relaxation, using Gurobi version 9. Then, using deterministic rounding, we obtain a $6$-approximate integral solution. Due to their large size, for the CoverType \cite{CovType} and the US Census \cite{USCensus} datasets, we split them into batches of size 1K, and for each batch, we compute a $6$-approximation to the offline optimal solution as above. 

Subsequently, for each demand point $v$, we generate the associated  prediction $p$ using the following approaches:

\begin{enumerate}
\item \verb|alpha_predictor|: $p$ is located on the line connecting $v$ to the optimal center $c^\ast$ where $v$ is assigned at distance $\alpha \, d(v,c^{\ast}_{v})$ to $c^\ast$.

%\item \verb|alpha_gaussian_predictor|: 
%$p$ is at distance $g_\alpha \, d(v,c^{*})$ to $c^{*}$, where $g_\alpha$ is drawn from a normal distribution $\mathcal{N}(\alpha, \textrm{std})$ with mean value $\alpha$ and standard deviation $\textrm{std}$.

\item \verb|gaussian_predictor|: $p$ is located on the line connecting $v$ to the optimal center $c^\ast$ at distance $g_\alpha \, d(v,c^{\ast}_{v})$ to $c^\ast$, where $g_\alpha$ is independently sampled from a normal distribution $\mathcal{N}(\alpha, \textrm{std})$ with mean value $\alpha$ and standard deviation $\textrm{std}$. We note that  \verb|gaussian_predictor| reduces to \verb|alpha_predictor| when \( \textrm{std} = 0 \).
\end{enumerate}

\begin{figure}[bht]
\centering
\includegraphics[width=.352\textwidth]{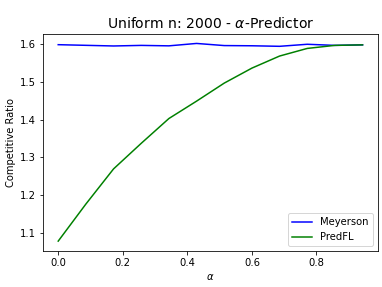} 
\includegraphics[width=.315\textwidth]{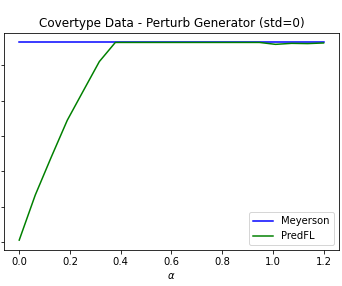} 
\includegraphics[width=.315\textwidth]{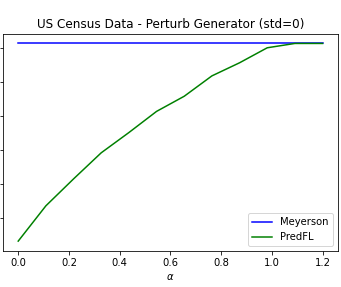} 
\caption{A comparative analysis of \nameref{alg:greedy3_fl}'s and \textsc{MeyOFL}'s competitive ratio across various prediction sequences and errors.}
\label{fig:experiments}
\end{figure}

%\smallskip
\noindent\textbf{Results.} 
In \cref{fig:experiments}, we plot the competitive ratio of \nameref{alg:greedy3_fl} against the competitive ratio of Meyerson's algorithm \cite{Meye2001} (\textsc{MeyOFL}) as a function of the parameter $\alpha$. Across all three datasets, we observe a consistent behavior: \nameref{alg:greedy3_fl}'s competitive ratio converges smoothly to that of \textsc{MeyOFL} as the parameter $\alpha$, which determines the prediction error, increases from $0$ to $1$. 

\begin{figure}
% \begin{multicols}{2}
\begin{subfigure}{.45\textwidth}
    \includegraphics[width=\linewidth]{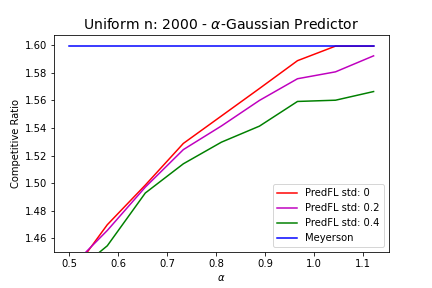}
    % \par 
    \includegraphics[width=\linewidth]{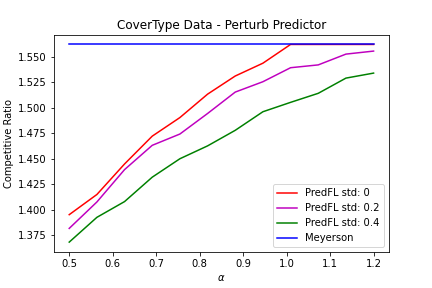}
    % \par 
    \includegraphics[width=\linewidth]{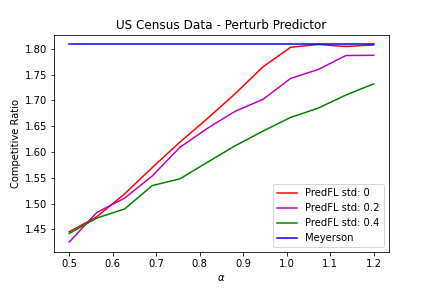}
    \caption{Competitive ratio of \nameref{alg:greedy3_fl} and \textsc{MeyOFL} across various std values.}
\label{fig:experiments_std}
\end{subfigure}
% \end{multicols}
% \begin{multicols}{2}
\begin{subfigure}{.45\textwidth}
    \includegraphics[width=\linewidth]{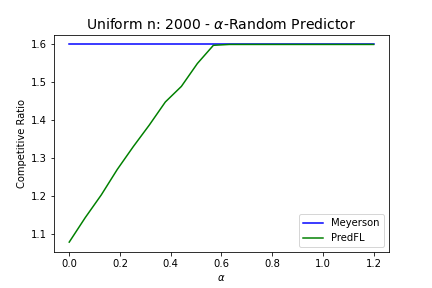}
    % \par 
    \includegraphics[width=\linewidth]{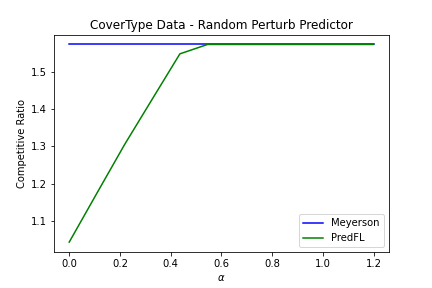}
    % \par 
    \includegraphics[width=\linewidth]{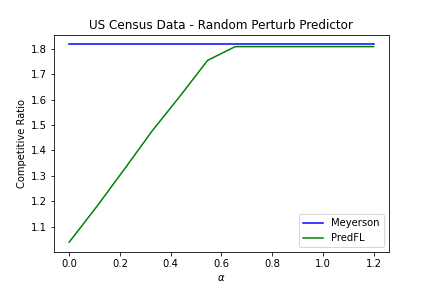}
    \caption{Competitive ratio of \nameref{alg:greedy3_fl} and \textsc{MeyOFL} for random reflections.}
\label{fig:experiments_random}
\end{subfigure}
% \end{multicols}
% \caption{caption here}
\end{figure}

%\smallskip\noindent{\bf Standard Deviation.}
To better understand the effect of the standard deviation of the prediction error on the performance of \nameref{alg:greedy3_fl}, we selected three representative \textrm{std} values and repeated the experiments for each value with the \verb|gaussian_predictor|, as shown in \cref{fig:experiments_std}. We truncated the value of $g_\alpha$ to be within the range $[0,1]$, because we wanted to ensure that each prediction always lies between the optimal center and the demand location. 
%
% As elucidated in \cref{sec:exper}, we observed a slight decrease in the competitive ratio with increasing standard deviation for both predictors, \verb|alpha_gaussian_predictor| and \verb|perturb_gaussian_predictor|.
% An interesting finding is that for \verb|alpha_gaussian_predictor|, the total cost of the \nameref{alg:greedy3_fl} is slightly decreasing with the standard deviation. On the one hand, increasing the standard deviation might cause $\eta_\infty$ to increase. On the other hand, some of the subsequent far predictions are bound to be closer to the optimal center, and due to our facility opening rule these predictions are more likely to turn into new facilities, due to their increased distance to the nearest open facility. Interestingly, the latter effect dominates, which results in a slightly decreased competitive ratio, as the standard deviation increases. This is in accordance with our lower bound construction, where $d(v, c_{open}) / d(p_v, c_{open}) = \alpha$ where $c_{open}$ is the, at the time, open facility opened by \nameref{alg:greedy3_fl} and $v$ a demand.
%
An interesting finding is that for the \verb|gaussian_predictor|, the competitive ratio of \nameref{alg:greedy3_fl} slightly improves as standard deviation increases. This finding presents a nuanced interplay of factors. On the one hand, increasing standard deviation may lead to an increase in $\eta_\infty$. On the other hand, the best predictions may be even closer to the optimal center. Due to our facility opening rule, the latter predictions are more likely to cause new facilities to open, due to their increased distance to the nearest algorithm's facility. Remarkably, the latter effect appears to dominate, resulting in a marginal decrease in the competitive ratio as \textrm{std} increases. %Interestingly, this aligns with our lower bound construction, where $d(v, c_{open}) / d(p_v, c_{open}) = \alpha$, with $c_{open}$ denoting the currently open facility opened by \nameref{alg:greedy3_fl} and $v$ representing a demand.

% An interesting finding of our experiments (the corresponding plots are deferred
% to the appendix) is that in \verb|alpha_gaussian_predictor|, the total cost of
% the online algorithm is slightly decreasing with the standard deviation. On the
% one hand, increasing the standard deviation might cause $\eta_\infty$ to
% increase. On the other hand, some of the subsequent far predictions are bound
% to be closer to the optimal center, and due to our facility opening rule these
% predictions are more likely to turn into new facilities, due to their increased
% distance to the nearest open facility. Interestingly, the latter effect
% dominates, which results in a slightly decreased competitive ratio, as the
% standard deviation increases. This is in accordance with our lower bound
% construction, where $d(x, f_{open}) / d(\hat{f}_x, f_{open}) = \alpha$, for all
% demands $x$.  

%\begin{figure}[!htp]
%
%\centering
%\includegraphics[width=.4\textwidth]{figures/uniform_alpha_cp}
%\includegraphics[width=.4\textwidth]{figures/covertype_avg_cp.png}  
%\includegraphics[width=.4\textwidth]{figures/uscensus_avg_cp.png} 
%\caption{Caption}
%\label{fig:experiments}
%
%\end{figure}

% \subsection{Random Reflections}
\smallskip\noindent{\bf Random Reflections.}
In \cref{fig:experiments_random}, we assess the competitive ratio of our algorithm using a third approach to prediction generation: %\verb|random_alpha_predictor| and 
\verb|random_perturb_predictor|. For each new demand $v$ mapped to an optimal center $c^{*}$, the associated prediction $p$ is located on the line connecting $v$ to $c^\ast$ at distance $\alpha \, d(v, c^{*})$ to $c^\ast$. However, the prediction is then randomly reflected across all possible perpendicular hyperplanes passing through $c^{*}$. This is done by multiplying the vector $p-c^{*}$ by a random $\pm 1$ vector of the same dimension as $p$. Consequently, the prediction is obtained by adding the resulting vector to $c^{*}$.

We observe note that across all datasets, the competitive ratio of \nameref{alg:greedy3_fl} matches that of Meyerson's algorithm as soon as the parameter $\alpha$ approaches $0.5$. This happens because random reflections may generate predictions with distances to the nearest algorithm's facility significantly larger than those of predictions without random reflections. Consequently, due to our facility opening rule, these predictions are more likely to cause a new facility to open, thereby leading to an increased competitive ratio. %Fortunately, this increase in competitive ratio is limited to, at most, a constant factor.

% \begin{figure}[!htb]
% \centering
% \includegraphics[width=.49\textwidth]{figures/random/uniform_random_cp.png}
% \includegraphics[width=.49\textwidth]{figures/random/covertype_random_cp.png}
% \includegraphics[width=.49\textwidth]{figures/random/uscensus_random_cp.png}
% \caption{A comparison of PredFL's and Meyerson's competitive ratio for random reflections.}
% \label{fig:experiments_random}
% %TODO edit the name of the third graph
% \end{figure}

% \input{conclusions}

%%%%%%%%%% 

%%%%%%%%%%%%%%%%%%%%%%%%%%%%%%%%%%%%%%%%%%%%%%%%%%%%%%%%%%%%

% Bibliography
% \bibliographystyle{unsrt}
%\bibliographystyle{splncs04}
\bibliographystyle{plain}

%\bibliography{short_references}
\bibliography{references2}

%%%%%%%%%%%%%%%%%%%%%%%%%%%%%%%%%%%%%%%%%%%%%%%%%%%%%%%%%%%%
\newpage
\appendix

\end{document}